\documentclass[twocolumn,aps,prx,amsmath,amssymb,superscriptaddress]{revtex4}
\usepackage{graphicx,dcolumn,bm,color,framed}
\usepackage{ulem}
\usepackage{bm}

\begin{document}

\title{Lorenz ratio of an impure compensated metal in the degenerate Fermi liquid regime}

\author{Woo-Ram Lee}
\affiliation{Department of Physics and Astronomy, The University of Alabama, Tuscaloosa, Alabama 35487, USA}
\affiliation{Department of Physics, Virginia Tech, Blacksburg, Virginia 24061, USA}
\author{Karen Michaeli}
\affiliation{Department of Condensed Matter Physics, The Weizmann Institute of Science, Rehovot 76100, Israel}
\author{Georg Schwiete}
\affiliation{Department of Physics and Astronomy, The University of Alabama, Tuscaloosa, Alabama 35487, USA}


\begin{abstract}
The Lorenz ratio serves as a measure to compare thermal and electric conductivities of metals. Recent experiments observed small Lorenz ratios in the compensated metal WP$_2$, indicating that charge flow is strongly favored over heat conduction. Motivated by these findings, we study transport properties of compensated metals in the presence of electron-electron collisions and  electron-impurity scattering. We focus on intermediate temperatures, where the phonon contributions to transport are weak and elastic and inelastic scattering rates are comparable. Our exact solution for the kinetic equation in the presence of general Fermi-liquid interactions is used  to extract the Lorenz ratio for short and long range interactions.  We find that the Lorenz ratio develops a temperature dependence and gets enhanced as a consequence of disorder scattering. For collisions mediated by the Coulomb interaction, impurities give rise to a non-monotonic dependence of the Lorenz ratio on the screening wave number with a minimum for intermediate screening strength. To help future experimental efforts, we establish a scheme to connect the exact results with the solution of the Boltzmann equation under the relaxation time approximation for all collision integrals. Our recipe provides simple phenomenological expressions for the transport coefficients and it allows for a physically transparent interpretation of the results. 

\end{abstract}

\maketitle

\section{Introduction}

Transport coefficients of  metals can significantly vary with temperature~\cite{Ziman01}. A primary source of temperature dependence is the various mechanisms of electron scattering, such as electron-phonon and electron-electron interactions.  Over a wide range of temperatures, the scattering of electrons by phonons is the most important means of relaxing the currents, and hence, it controls the transport coefficients. When the temperature decreases, all inelastic scattering rates are suppressed  and the residual resistance observed at the lowest temperatures is caused by the scattering of electrons on impurities.  Electron-electron collisions can become important at intermediate temperatures.  In many cases, electron-electron collisions alone cannot contribute  to the electric resistance, since they conserve the total momentum of the participating particles and prevent the charge current from decaying.  Notable counterexamples arise in electronic systems where the underlying lattice plays an important role, and umklapp scattering relaxes the current.   Another interesting case occurs in compensated metals, which have both electron and hole bands and equal population of charge carriers. In these systems, the current is proportional to the \textit{difference} between the total electron and the total hole momenta, and thus, can relax by collisions between the two types of particles. This mechanism was discussed by Baber in Ref.~\cite{Baber37}. 

Transport properties of compensated metals such as Mg, Zn, Cd, Bi, and W (for a more comprehensive list, see \cite{Fawcett63}) have already been well-known for a long time. Recent experiments on WP$_2$ \cite{Jaoui18, Gooth18}, however, sparked  renewed interest in their low-temperature transport properties \cite{Li18}. These experiments studied the temperature dependence of  the ratio between the thermal ($\kappa$) and electric ($\sigma$) conductivities,  i.e., the Lorenz ratio $\mathcal{L}=\kappa/T\sigma$. In the $T\rightarrow 0$ limit, metals feature a universal ratio $\mathcal{L}_0=\pi^2/(3e^2)$ with $-e$ the electron charge~\cite{units}. The ratio measured in Ref.~\cite{Jaoui18, Gooth18} has a minimum, $\mathcal{L}\lesssim 0.25\mathcal{L}_0$, at  a low but finite temperature $T_{m}\sim 10$ K.  A Lorenz ratio smaller than unity implies that charge transport is more efficient than heat conduction. Analysis of the experimental result~\cite{Jaoui18} concluded that the effect of electron-phonon scattering on the transport coefficients in WP$_2$ is negligible at $T\sim T_m$. Thus, the  smallness of the Lorenz ratio is likely caused by electron-electron interactions.

Previous  efforts  to analyze the effect of electron-electron collisions on the electric and thermal transport properties of WP$_2$ assumed a clean, compensated metal~\cite{Li18} . The small value of the Lorenz ratio was then attributed to weakly screened Coulomb interactions that give rise to small-angle scattering. Such a mechanism is  effective in relaxing the flow of heat  but not of charge.  An alternative origin for a small Lorenz ratio could be strong intra-band scattering, which predominantly affects the thermal conductivity. The model of Ref.~\cite{Li18} provides valuable insight into the influence of different electron-electron scattering mechanisms on the Lorenz ratio. It does not, however, explain its temperature dependence.  At low temperatures, all scattering rates are proportional to $T^2$ and therefore, the  temperature drops out from the Lorenz ratio $\kappa/(T\sigma)$.

Motivated by the experiments on WP$_2$ \cite{Jaoui18, Gooth18}, we study transport properties of general compensated metals with non-vanishing electron and hole densities. Our work focuses on the combined effect of electron-electron collisions (both interband and intraband) and electron-impurity scattering on the temperature dependence of the Lorenz ratio. The inclusion of disorder ensures that the Lorenz ratio becomes $\mathcal{L}_0$ as $T\rightarrow 0$  (as observed in the experiment \cite{Jaoui18}).   We consider two types  of inelastic collisions: (i) Hu bbard-like electron-electron interactions that scatter electrons uniformly and (ii) screened Coulomb interactions, for which small-angle scattering is dominant. We find that for both types of interactions, the Lorenz number  decreases with increasing temperature, in agreement with the experimental result.  

The most significant reduction in $\mathcal{L}(T)$ occurs up to temperatures where the elastic and inelastic collisions rates become equal. Interestingly, a minimum in the Lorenz ratio can be caused by electron-electron interactions alone. Moreover, our result indicates that the minimal value of the Lorenz ratio  in real systems is limited by the disorder strength. For weak disorder, electron-electron and electron-impurity scattering rates can become comparable at temperatures where phonons are still negligible. Then, the Lorenz ratio can take values significantly below $\mathcal{L}_0$, before it increases with temperature due to phonon contributions to the heat flow. Thus, the experimental observations \cite{Jaoui18, Gooth18} at low temperatures can be explained by the interplay of electron-electron and electron-impurity scattering. By contrast, in highly disordered systems, phonons are expected to change the Lorenz ratio well before electron-electron interaction effects become relevant \cite{Remark6}. 

We derive the transport coefficients from the kinetic equation describing a multi-band system with both electrons and holes. To keep the model tractable, we choose a spherical Fermi surface with a quadratic dispersion for both charge carriers. In addition, we introduce two types of  independent collision integrals that describe the effect of elastic and inelastic scattering. We first simplify the kinetic equation by using the relaxation-time approximation (RTA) for all collisions. This step allows us to obtain an exact solution for all transport coefficients, as demonstrated in Ref.~\cite{Keyes58} for the electric conductivity of a single-band metal, and recently been applied to other transport coefficients in Ref.~\cite{Lee20}. A similar method has been used for calculating the electric conductivity for a two-band compensated metal~\cite{Gantmakher78, Kukkonen79}; we generalize those works to also compute the thermal conductivity and Lorenz ratio. To account for the details of the inelastic interactions, we also solve the full kinetic equation for Fermi-liquid like interactions. For this purpose, we employ a mapping of the linearized Boltzmann equation to a second order differential equation of the Schr\"odinger type, which can be solved exactly by an eigenfunction expansion. This approach was previously used to study transport properties of metals with~\cite{Bennett69,Smith69} and without~\cite{Jensen68,Brooker68,Sykes70} elastic scattering. The same method was applied to calculate the Lorenz ratio of compensated metals in the clean limit~\cite{Li18} and the electric conductivity of compensated metals for special values of the scattering rates~\cite{Kukkonen79}. Recently, we derived a set of formulas for the electric and thermal conductivities of a disordered one-band model \cite{Lee20}, which allows  for a straightforward numerical evaluation. Here, we derive analogous expressions for the compensated metal and analyze the results.

A comparison between the two solutions reveals that a simple RTA cannot capture the different effect that interband scattering has on electric and heat transport: While forward scattering events  reduce the thermal conductivity they barely change the electric conductivity. We show that the RTA can be easily modified to reproduce the correct temperature dependence of all transport coefficients.  We achieve this by introducing (phenomenologically) different interband relaxation times in the solution of the charge and heat conductivities.   Thus, we provide a tractable and simple solution for the transport coefficients  as well as the Lorenz ratio of compensated metals in the presence of both disorder and electron-electron interactions.

This manuscript is structured as follows. In Sec.~\ref{sec:LinBoltz}, we introduce our model of a compensated metal and discuss the structure of the linearized Boltzmann equation. In Sec.~\ref{sec:RTA}, we introduce a simplified kinetic equation, where all collision integrals are treated in the RTA. We derive closed-form expressions for the electric, thermal and thermoelectric transport coefficients and the Lorenz ratio. In Sec.~\ref{sec:FLCI}, we use the method of Refs.~\cite{Jensen68,Brooker68,Bennett69,Sykes70,Lee20} to solve the Boltzmann equation and calculate the conductivities of  impure compensated metals with Fermi-liquid collision integrals for the intraband and the interband scattering. In Sec.~\ref{sec:discussion}, we use two specific model interactions, Hubbard-type short-range interaction and statically screened Coulomb interaction, to discuss various aspects of our result for the transport coefficients. Moreover, we compare the exact solution of the kinetic equation with  the results obtained using the RTA. We conclude in Sec.~\ref{sec:conclusion}.


\section{Linearized Boltzmann equation for the compensated metal}
\label{sec:LinBoltz}

\begin{figure}[t]
\centering
\includegraphics[width=0.45\textwidth]
{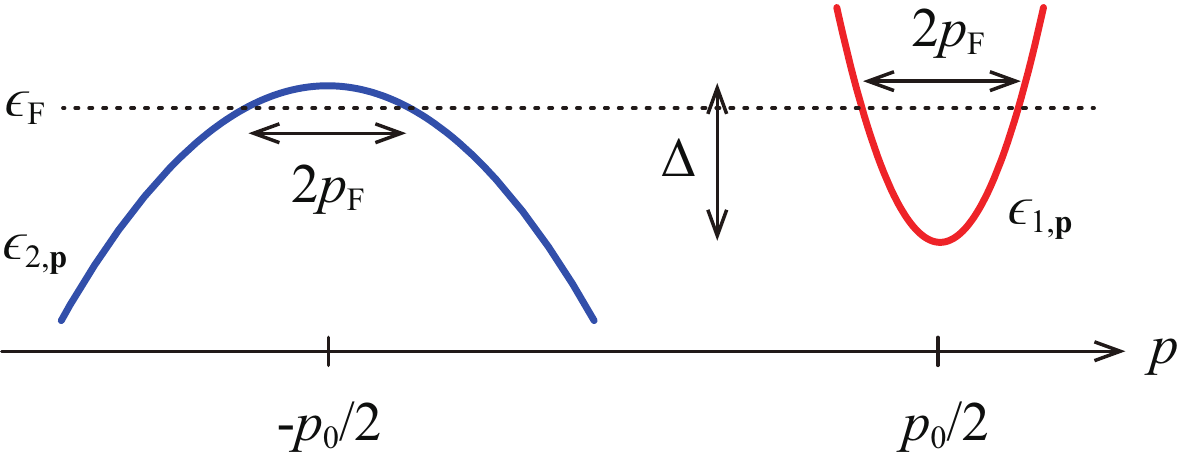} \\
\caption{
The electron and hole bands for the compensated metal studied in this manuscript.
}
\label{Figure0}
\end{figure}

A compensated metal (CM) is a metal with equal numbers of electrons and holes. Here, we study a low-energy model described by the combination of electron and hole bands (see Fig.~\ref{Figure0}) with quadratic dispersion \cite{Baber37,Li18} 
\begin{align}
\epsilon_{1,{\bf p}}
& = \frac{({\bf p} - {\bf p}_0/2)^2}{2m_1},
\\
\epsilon_{2,{\bf p}}
& = - \frac{({\bf p} + {\bf p}_0/2)^2}{2m_2} + \Delta.
\label{eq:model}
\end{align}
In this equation, $\Delta$ is the energy offset, and $m_1$ ($m_2$) is the electron (hole) effective mass. If we set the electron and hole densities to be $\mathcal{N}_1 = \mathcal{N}_2 = \mathcal{N}$, the Fermi momentum and Fermi energy are given by $p_F = (3\pi^2\mathcal{N})^{1/3}$ and $\epsilon_F = \Delta m_2 / (m_1 + m_2)$, respectively. 

We use the Boltzmann equation \cite{Ziman01} to study electric and thermal transport in a compensated metal. For the electron band, the collision integral consists of three parts: $I_1\{f_1,f_2\} = I_{ei}^{11}\{f_1\} + I_{ee}^{11}\{f_1\} + I_{ee}^{12}\{f_1,f_2\}$. Here, $I_{ei}^{11}$, $I_{ee}^{11}$, and $I_{ee}^{12}$ stand for intraband electron-impurity scattering (with the interband counterpart ignored for simplicity), intraband and interband electron-electron scattering, respectively, and $f_1$ $(f_2)$ is the distribution function for the electron (hole) band. The collision integral for the hole band $I_{2}\{f_1,f_2\}$ is obtained by switching the indices $1 \leftrightarrow 2$. We use the RTA for electron-impurity scattering: the distribution function $f_c({\bf r},{\bf p},t)$ relaxes towards the angularly averaged distribution $\langle f_c({\bf r},{\bf p},t)\rangle$ in a characteristic time $\tau_{ei,c}$,
\begin{align}
I_{ei}^{cc}\{f\}
= - \frac{f_{c}({\bf r},{\bf p},t) - \langle f_{c}({\bf r},{\bf p},t) \rangle}{\tau_{ei,c}}.
\label{Coll_ei}
\end{align}
Hereafter, it is understood that momenta in the electron/hole bands are measured from the points $\pm {\bf p}_0/2$, respectively.

Our aim is to find the transport coefficients in linear response to either an electric field ${\bf E}$ or a temperature gradient $\nabla_{\bf r} T$. To this end, the distribution function can be expanded as $f_c({\bf r}, {\bf p}) \approx n_F(\xi_{c,{\bf p}}) + \delta f_{c,{\bf p}}$. Here, $n_F(\xi_{c,{\bf p}}) = [\exp(\beta\xi_{c,{\bf p}}) + 1]^{-1}$ is the Fermi-Dirac distribution with $\beta = 1/T$, and $\xi_{c,{\bf p}} = \epsilon_{c,{\bf p}} - \mu$. 

The linearized Boltzmann equations for electrons and holes are coupled by the interband scattering processes, 
\begin{align}
& ~~~ \left( -e{\bf E} - \xi_{c,{\bf p}} \frac{\nabla_{\bf r}T}{T} \right) \cdot \boldsymbol{v}_{c,{\bf p}} \frac{\partial n_F(\xi_{c,{\bf p}})}{\partial\xi_{c,{\bf p}}}
\nonumber\\
& = - \frac{\delta f_{c,{\bf p}}}{\tau_{ei,c}} 
+ I_{ee}^{cc}\{\delta f\} + I_{ee}^{c\bar{c}}\{\delta f\}.
\label{BoltzmannEq_Linearized}
\end{align}
where we denote $\bar{c}=2$ for $c=1$ and vice versa. The group velocities of the electron and hole bands are defined by $\boldsymbol{v}_{1,{\bf p}} = \nabla_{\bf p} \epsilon_{1,{\bf p}} = {\bf p}/m_1$ and $\boldsymbol{v}_{2,{\bf p}} = \nabla_{\bf p} \epsilon_{2,{\bf p}} = - {\bf p}/m_2$, respectively. The electron-electron collision integrals are also linearized.

\section{Relaxation-time approximation}
\label{sec:RTA}
Closed-form solutions to the coupled set of Boltzmann equations \eqref{BoltzmannEq_Linearized} can be derived for a model in which {\it all} collision integrals are treated in the RTA. In this section, we introduce such a model by generalizing the single-band Boltzmann equation first discussed by Keyes \cite{Keyes58} to a compensated metal with two bands. This allows us to find the distribution function in linear response to an electric field or a thermal gradient, and to calculate the electric, thermal and thermoelectric transport coefficients. While individual scattering processes are not described accurately in the RTA, the main merit of this approach is that it provides valuable insight into the {\it interplay} of the three scattering processes and how they affect the different transport coefficients.

The intraband electron-electron collision integral $I_{ee}^{cc}$ in the RTA can be set up in analogy to the single-band case \cite{Keyes58,Lee20},
\begin{align}
I_{ee}^{cc}\{f_c\}
= - \frac{f_c({\bf r},{\bf p}) - n_{F,c}^{({\rm c.m.})}(\bf p)}{\tau_{ee}^{cc}}.
\label{Coll_ee}
\end{align}
Here, $\tau_{ee}^{cc}$ is the characteristic time for intraband electron-electron scattering and c.m. stands for ``center of mass". We also introduced the ``drifting" Fermi-Dirac distribution function $n_{{\rm F},c}^{({\rm c.m.})}$. This distribution function is related to the the Fermi-Dirac distribution in the laboratory frame (in which impurities are at rest) as
\begin{align}
n_{{\rm F},c}^{({\rm c.m.})}({\bf p}) 
= n_{{\rm F},c}(\epsilon_{c,{\bf p}} - \mu - {\boldsymbol v}_c^{({\rm c.m.})} \cdot {\bf p}).
\label{eq:ncm}
\end{align}
For a quadratic dispersion, the drift velocity ${\boldsymbol v}_c^{({\rm c.m.})}$ corresponds to the velocity of the center-of-mass motion of electrons and holes in their respective bands \cite{Remark4}, 
\begin{align}
{\boldsymbol v}_c^{({\rm c.m.})} = s \int_{\bf p} {\boldsymbol v}_{c,{\bf p}} \delta f_{c,{\bf p}} / \mathcal{N}. \label{eq:velocity}
\end{align}
Here, $s=2$ is the spin degeneracy and particle and hole densities are equal and denoted by the common symbol $\mathcal{N} = s \int_{\bf p} n_F(\xi_{1,{\bf p}}) = s \int_{\bf p} [1 - n_F(\xi_{2,{\bf p}})]$.

A finite drift velocity arises only when the system is taken out of equilibrium. This is why in linear response one may expand $n_{F,c}^{({\rm c.m.})}$ to first order in ${\boldsymbol v}_c^{({\rm c.m.})}$. In this way, one obtains the expression for the intraband collision integral that will be used for the calculations presented in this section,
\begin{align}
I_{ee}^{cc}\{f\} = - \frac{1}{\tau_{ee}^{cc}} \bigg[ \delta f_{c,{\bf p}} + {\eta_c} {\boldsymbol v}_c^{({\rm c.m.})} \cdot {\bf p} \frac{\partial n_{F,c}(\xi_{c,{\bf p}})}{\partial\xi_{c,{\bf p}}} \bigg].
\label{I_ee_11}
\end{align}
Here, we used the notation $\eta_1=-\eta_2=1$. 

The interband electron-electron collision integral can be formulated in a similar spirit
\begin{align}
I_{ee}^{c\bar{c}}\{f\} = - \frac{1}{\tau_{ee}^{c\bar{c}}} \bigg[ \delta f_{c,{\bf p}} - {\eta_c} {\boldsymbol v}_{\bar{c}}^{({\rm c.m.})} \cdot {\bf p}\frac{\partial n_{F,c}(\xi_{c,{\bf p}})}{\partial\xi_{c,{\bf p}}} \bigg],
\label{I_ee_12}
\end{align}
where $\tau_{ee}^{c\bar{c}}$ is the characteristic time for interband electron-electron collisions. 

There is an important difference between intraband and interband collisions. Intraband collisions conserve the momentum for each band separately, $\int_{\bf p} {\bf p}\; I_{ee}^{cc}\{f\}=0$. Interband collisions, in turn, conserve the total momentum for the combined system consisting of particles and holes, $\int_{\bf p} {\bf p}  \left(I_{ee}^{12}\{f\}+I_{ee}^{21}\{f\}\right)=0$. This condition imposes an additional constraint on the parameters characterizing the interband collision integral, namely $m_1/\tau_{ee}^{12}=m_2/\tau_{ee}^{21}$. This observation suggests introducing a new timescale $\tau$ via the relation 
\begin{align}
\frac{m_c}{\tau_{ee}^{c\bar{c}}} = \frac{\tilde{m}}{\tau},
\end{align}
where $1/\tilde{m} = 1/m_1 + 1/m_2$. The time $\tau$ is the characteristic time scale for the decay of the electric current in the system consisting of particles and holes, as identified in \cite{Gantmakher78}. Reference~\cite{Gantmakher78} studied the electric conductivity of a compensated metal with impurities and interband scattering using an equation of motion approach. These equations of motion can be obtained from the set of coupled Boltzmann equations by calculating the average momenta for the two bands.

\subsection{Non-equilibrium distribution}
\label{Subsec:NED}

The coupled Boltzmann equations, Eq.~\eqref{BoltzmannEq_Linearized}, for $c\in \{1,2\}$ with the collision integral $I_{ee}$ given in Eqs.~\eqref{I_ee_11} and \eqref{I_ee_12} can formally be resolved for $\delta f_c$ as
\begin{align}
\delta f_{c,{\bf p}} 
& = \tilde{\tau}_c \boldsymbol{v}_{c,{\bf p}} \cdot \bigg( e \tilde{{\bf E}}_c + \xi_{c,{\bf p}} \frac{\nabla_{\bf r} T}{T} \bigg) \frac{\partial n_{F,c}(\xi_{c,{\bf p}})}{\partial \xi_{c,{\bf p}}},
\label{TrialSolution}
\end{align}
with the effective electric field 
\begin{align}
\tilde{{\bf E}}_c 
= {\bf E} - \frac{m_c {\boldsymbol v}_c^{({\rm c.m.})}}{e\tau_{ee}^{cc}} + \frac{m_c {\boldsymbol v}_{\bar{c}}^{({\rm c.m.})}}{e\tau_{ee}^{c\bar{c}}},
\label{EffectiveElectricField}
\end{align}
and the total scattering rate
\begin{align}
\frac{1}{\tilde{\tau}_c}
= \frac{1}{\tau_{ei,c}} + \frac{1}{\tau_{ee}^{cc}} + \frac{1}{\tau_{ee}^{c\bar{c}}}=\frac{1}{\tau_{ei,c}} + \frac{1}{\tau_{ee}^{cc}} + \frac{\tilde{m}}{m_c\tau}
\label{TotalScatteringRate}
\end{align}
One more step is necessary in order to find $\delta f_c$, because the right-hand side of Eq.~\eqref{TrialSolution} depends on $\delta f_c$ implicitly through ${\boldsymbol v}_c^{\rm (c.m.)}$. To make progress, we follow the idea outlined in Ref.~\cite{Keyes58} and insert the formal solution for $\delta f_c$ in Eq.~\eqref{TrialSolution} into the definition of the center of mass velocity \eqref{eq:velocity}. In this way, one obtains a coupled set of equations for the center of mass velocities,
\begin{align}
& ~~~ \left(
\begin{array}{cc}
m_1 (1/\tau_{ei,1} + 1/\tau_{ee}^{12}) & m_1/\tau_{ee}^{12}
\\
m_2/\tau_{ee}^{21} & m_2 (1/\tau_{ei,2} + 1/\tau_{ee}^{21})
\end{array}
\right) \left(
\begin{array}{c}
{\boldsymbol v}_1^{({\rm c.m.})}
\\
{\boldsymbol v}_2^{({\rm c.m.})}
\end{array}
\right)
\nonumber\\
& = \left(
\begin{array}{c}
- e{\bf E} - \langle\!\langle \xi_{1,{\bf p}} \rangle\!\rangle \nabla_{\bf r} T/T
\\
- e{\bf E} - \langle\!\langle \xi_{2,{\bf p}} \rangle\!\rangle \nabla_{\bf r} T/T
\end{array}
\right).\label{eq:vKeyes}
\end{align}
where $\left\langle\!\left\langle\dots\right\rangle\!\right\rangle$ denotes the average
\begin{align}
\langle\!\langle X_{c,{\bf p}} \rangle\!\rangle
& = - \frac{s m_c}{d\mathcal{N}} \int_{\bf p} X_{\bf p} v_{c,{\bf p}}^2 \frac{\partial n_{F}(\xi_{c,{\bf p}})}{\partial\xi_{c,{\bf p}}},
\label{FluctuationAverage}
\end{align}
This result for ${\boldsymbol v}_c^{({\rm c.m.})}$ can be inserted into Eq.~\eqref{EffectiveElectricField} to find the non-equilibrium part of the distribution functions, 
\begin{align}
\delta f_{c,{\bf p}}^E
& = \frac{\tau_{ei,c} \boldsymbol{v}_{c,{\bf p}} \cdot e{\bf E}}{1 + (\tilde{m}/\tau)(\tau_{ei,1}/m_1 + \tau_{ei,2}/m_2)} \frac{\partial n_{F,c}(\xi_{c,{\bf p}})}{\partial \xi_{c,{\bf p}}},
\label{DistFluc_E}
\\
\delta f_{c,{\bf p}}^T
& = \left[ \frac{\displaystyle \tau_{ei,c} \bigg\{ \langle\!\langle \xi_{c,{\bf p}} \rangle\!\rangle + \frac{\tilde{m}}{\tau} \frac{\tau_{ei,\bar{c}}}{m_{\bar{c}}} (\langle\!\langle \xi_{c,{\bf p}} \rangle\!\rangle - \langle\!\langle \xi_{\bar{c},{\bf p}} \rangle\!\rangle) \bigg\}}{\displaystyle 1 + \frac{\tilde{m}}{\tau} \bigg( \frac{\tau_{ei,1}}{m_1} + \frac{\tau_{ei,2}}{m_2} \bigg)} \right.
\nonumber\\
& ~~~ + \tilde{\tau}_c (\xi_{c,{\bf p}} - \langle\!\langle \xi_{c,{\bf p}} \rangle\!\rangle) \bigg] \boldsymbol{v}_{c,{\bf p}} \cdot \frac{\nabla_{\bf r} T}{T} \frac{\partial n_{F,c}(\xi_{c,{\bf p}})}{\partial \xi_{c,{\bf p}}}.
\label{DistFluc_T}
\end{align}

The non-equilibrium part of the distribution functions, Eqs.~\eqref{DistFluc_E} and \eqref{DistFluc_T}, have been derived for a momentum-independent electron-impurity scattering rate. A generalization to include the momentum dependence of $\tau_{ei,c}$ is straightforward within the formalism presented in this section, but this discussion is beyond the scope of this manuscript.

In the introduction of Sec.~\ref{sec:RTA}, we discussed the properties of the electron-electron collision integral. We focus here on the constraints imposed on the relaxation times to guarantee total energy and momentum conservation. To explain the consequences of momentum conservation on the transport properties of compensated metals, we start with comparing the results given by Eqs. \eqref{DistFluc_E} and \eqref{DistFluc_T} to that of a single-band~\cite{Keyes58,Lee20}. The single-band case can be recovered by switching off the interband scattering, $\tau\rightarrow \infty$, for a fixed $c$. In the absence of any momentum relaxation mechanism, the force created by an electric field accelerates the particles in the band irrespective of the intraband interactions. Thus, the total momentum of the corresponding Fermi sea grows at a constant rate, and the electric conductivity diverges. This mechanism manifests itself in the expression for $\delta f^E$ of a single band: it depends on $\tau_{ei}$ but not on $\tau_{ee}$. For the compensated metal,  Eq.~\eqref{DistFluc_E} reveals that the intraband scattering rate also drops from the result for $\delta f^E_{c,{\bf p}}$. Another common feature of the single-band and compensated metals is the rigid, i.e., momentum-independent, shift of the Fermi surface(s) by the electric field. In the absence of interband scattering, the shift of the Fermi surfaces is determined by the elastic scattering rate alone. Interband scattering, in turn, leads to a dressing of the elastic scattering rate, which becomes temperature-dependent (via $\tau$) and is also influenced by the elastic scattering rate of the other band.

A temperature gradient gives rise to an energy-dependent  ($\xi$) force acting on particles and holes, Eq.~\eqref{TrialSolution}. The corresponding distribution function is given by Eq.~\eqref{DistFluc_T}.  
For a single band ($\tau\rightarrow \infty$), the non-equilibrium part of the distribution function is determined by two types of contributions: (i) a term depending on the average energy (measured from the Fermi energy) and (ii) a term proportional to the deviation of the energy from the average one. The former contribution can be interpreted as the response to a uniform force. Similar to the effect of an electric field, the resulting acceleration of the Fermi sea is limited by the relaxation of the total momentum alone. The latter term describes the response to a momentum-dependent force. The corresponding distortion of the Fermi surface is sensitive to all scattering mechanisms, and electron-electron interactions alone are sufficient to establish a finite shift. The absence of a stationary solution for the non-equilibrium distribution function due to the uniform force (i) in clean systems implies that a temperature gradient can drive a diverging thermal current. However, thermal conductivity is conventionally defined under the condition of a vanishing electric current, i.e., it is assumed that a counteracting electric field develops at the boundaries and cancels the uniform force. Therefore, the thermal conductivity of a single-band metal can be finite even in the clean limit. For the compensated metal, the acceleration of the Fermi seas by the uniform force cannot be stopped as long as the total momentum is conserved. Mathematically, this property is related to the divergence of the first contribution to $\delta f_{c,{\bf p}}^T$ given by Eq.~\eqref{DistFluc_T} in the clean limit $\tau_{ei,c}\rightarrow \infty$. The crucial distinction from the single-band case is that the condition of vanishing electric current is not sufficient to eliminate this diverging contribution to the thermal conductivity of compensated metals. This point will be discussed further in the next subsection.

A few remarks are in order here concerning Eq.~\eqref{eq:vKeyes} for the center of mass velocities. A peculiar feature of \eqref{DistFluc_E} is that the clean limit is not unambiguously defined (the order of $\tau_{ei,1}\rightarrow \infty$ and $\tau_{ei,2}\rightarrow \infty$ matters). Tracing back the origin of this in ambiguity, we see that the matrix in Eq.~\eqref{eq:vKeyes} becomes singular in the clean limit. Further inspection shows that for electric field-driving in the clean system, one may find the sum of the two center of mass velocities from this equation but not the difference. It is therefore possible to calculate the current and find the conductivity for the clean system, but not the total momentum or the individual distribution functions for the electron and hole band. For the compensated metal studied here, an electric field does not influence the total momentum of the electron-hole system, even though it can give rise to a current. Moreover, without impurities, none of the scattering mechanisms under consideration is able to influence the total momentum, which is therefore stationary. In the presence of impurities, the stationarity condition imposes an additional constraint $\int_{\bf p} \left(I_{ei,1}+I_{ei,2}\right)=0$ \cite{Gantmakher78}, which fixes the ratio of the center of mass velocities of the two bands and lifts the ambiguity. The same ambiguity arises for the clean limit when a temperature gradient is applied to the compensated metal. Here, the situation is slightly different compared to the electric-field driving: the temperature gradient can transfer a momentum to the electron-hole system. Without impurity scattering, the total momentum cannot relax and the system cannot be stationary. The consequences is discussed further below Eq.~\eqref{ThermalConductivity}. 

\subsection{Transport coefficients}
\label{Subsec:conductivities}

To find the transport coefficients, we insert the results for the non-equilibrium distribution functions Eqs.~\eqref{DistFluc_E} and \eqref{DistFluc_T} into the expressions for the electric and thermal current densities 
\begin{align}
\left(
\begin{array}{c}
{\bf J}_E
\\
{\bf J}_T
\end{array}
\right)
& = s \sum_{c\in\{1,2\}}\int_{\bf p}
\left(
\begin{array}{c}
-e
\\
\xi_{c,{\bf p}}
\end{array}
\right)
\boldsymbol{v}_{c,{\bf p}} \delta f_{c,{\bf p}}
\nonumber\\
& = \left(
\begin{array}{cc}
L_{EE} & L_{ET}
\\
L_{TE} & L_{TT}
\end{array}
\right) \left(
\begin{array}{c}
{\bf E}
\\
-\nabla_{\bf r} T
\end{array}
\right),
\label{CurrentDensity}
\end{align}
where $s=2$ appears due to the spin degeneracy. To formulate the results in a transparent form, it is useful to define an effective elastic scattering time in the following way
\begin{align}
\frac{\tau_{el}}{\tilde{m}}=\frac{\tau_{ei,1}}{m_1}+\frac{\tau_{ei,2}}{m_2}.\label{eq:tauel}
\end{align}
Note that if masses are comparable, and scattering times $\tau_{ei,c}$ unequal, $\tau_{el}$ is mainly determined by the larger  time.

Then, we can write the electric conductivity as
\begin{align}
\sigma
\equiv L_{EE}
=\frac{\mathcal{N}e^2}{\tilde{m}}\left(\frac{1}{\tau_{el}}+\frac{1}{\tau}\right)^{-1},
\label{Sigma_EE}
\end{align}
consistent with Ref.~\cite{Gantmakher78}. 

The thermoelectric transport coefficients satisfy the Onsager reciprocal relation, 
\begin{align}
L_{ET}
& = \frac{L_{TE}}{T}
= - \frac{\mathcal{N}e}{T} \frac{\displaystyle \frac{\tau_{ei,1}}{m_1} \langle\!\langle \xi_{1,{\bf p}} \rangle\!\rangle + \frac{\tau_{ei,2}}{m_2} \langle\!\langle \xi_{2,{\bf p}} \rangle\!\rangle}{\displaystyle 1 + \frac{\tau_{el}}{\tau} }.
\label{Sigma_ET}
\end{align}

For the thermal conductivity one can distinguish two cases. In open systems, where an electric current can flow, one obtains
\begin{align}
L_{TT}
& = \frac{\mathcal{N}}{T} \sum_{c\in\{1,2\}} \bigg[ \frac{\tilde{\tau}_c}{m_c} (\langle\!\langle \xi_{c,{\bf p}}^2 \rangle\!\rangle - \langle\!\langle \xi_{c,{\bf p}} \rangle\!\rangle^2) 
\nonumber\\
& ~~~ \left. + \frac{\tau_{ei,c}}{m_c} \frac{\displaystyle \langle\!\langle \xi_{c,{\bf p}} \rangle\!\rangle^2 + \frac{\tilde{m}}{\tau} \frac{\tau_{ei,\bar{c}}}{m_{\bar{c}}} \langle\!\langle \xi_{c,{\bf p}} \rangle\!\rangle (\langle\!\langle \xi_{c,{\bf p}} \rangle\!\rangle - \langle\!\langle \xi_{\bar{c},{\bf p}} \rangle\!\rangle)}{\displaystyle 1 + \frac{\tau_{el}}{\tau}} \right].
\label{Sigma_TT}
\end{align}
In experiment, one typically measures the thermal conductivity in the absence of electric current, ${\bf J}_E = 0$, for which one obtains \begin{align}
\kappa
& \equiv L_{TT} - \frac{L_{TE} L_{ET}}{L_{EE}}
\nonumber\\
& = \frac{\mathcal{N}}{T} \bigg[ \frac{\tilde{\tau}_1}{m_1} (\langle\!\langle \xi_{1,{\bf p}}^2 \rangle\!\rangle - \langle\!\langle \xi_{1,{\bf p}} \rangle\!\rangle^2) + \frac{\tilde{\tau}_2}{m_2} (\langle\!\langle \xi_{2,{\bf p}}^2 \rangle\!\rangle - \langle\!\langle \xi_{2,{\bf p}} \rangle\!\rangle^2) 
\nonumber\\
& ~~~ +\frac{(\langle\!\langle \xi_{1,{\bf p}} \rangle\!\rangle - \langle\!\langle \xi_{2,{\bf p}} \rangle\!\rangle)^2}{\frac{m_1}{\tau_{ei,1}}+\frac{m_2}{\tau_{ei,2}}} \bigg]\label{ThermalConductivity}
\end{align}

As anticipated, the electric conductivity in the clean case can be obtained straightforwardly by taking the limit $\tau_{ei,c}\rightarrow \infty$. The result takes a Drude-like form, $\sigma=\mathcal{N}e^2\tau/\tilde{m}$, where $\tau$ and $\tilde{m}$ play the role of effective scattering time and mass, respectively. Interband scattering results in a finite conductivity, even in the absence of elastic scattering. Intraband scattering, in turn, does not influence the electric conductivity, because it cannot relax the momenta of the bands and therefore does not lead to a decay of the current carried by each band. 

The result for $L_{ET}$ can be understood as a straightforward generalization of the single-band result \cite{Lee20}, albeit with an effective total scattering rate with elastic and inelastic (interband) contributions. This property is inherited from the form of $\delta f^E$, Eq.~\eqref{DistFluc_E}. 

The result for the thermal conductivity naturally falls into two parts, $\kappa=\kappa_a+\kappa_b$, corresponding to the first and second line on the right hand side of Eq.~\eqref{ThermalConductivity}. The expression for $\kappa_a$ is a straightforward generalization of the single-band case. Unlike for the electric conductivity, {\it all} scattering processes are effective: besides electron-impurity and interband scattering, intraband scattering also contributes to the thermal conductivity. The scattering rates simply add up to give $\tilde{\tau}_c$. The second contribution to the thermal conductivity, $\kappa_b$, is special in that it diverges in the clean limit. The reason has already been discussed in connection with $\delta f^T$ below Eq.~\eqref{DistFluc_T}. The thermal gradient gives rise to a net force on the electron-hole system, but without impurities none of the remaining scattering processes is able to keep the system in a stationary state. The same situation occurs in the single-band case, and the thermal gradient can induce a diverging charge current. However, thermal conductivity measurements are conventionally done under the condition of zero charge current, ${\bf J}_E=0$. Consequently, for a single band the diverging contribution is omitted from the thermal conductivity $\kappa$. Unlike in the single-band case, however, the zero-current boundary condition does not cure the divergency of the thermal conductivity for the compensated metal. The reason is that ${\bf J}_E=0$ does not imply a vanishing of the total momentum: electrons and holes moving with the same velocities do not produce a net charge current. Thus, the thermal conductivity retains the diverging contribution better known as the ambipolar (or bipolar) effect~\cite{Davydov40}. From Eq.~\eqref{ThermalConductivity}, we can see that in the presence of disorder the ambipolar contribution, $\kappa_b\sim \mathcal{N}\frac{T^3}{\epsilon_F^2}(\frac{m_1}{\tau_{ei,1}}+\frac{m_2}{\tau_{ei,2}})^{-1}$, no longer diverges. At sufficiently low temperatures, the ambipolar effect is negligible in comparison to $\kappa_a\sim \mathcal{N} T\left(\frac{\tilde{\tau}_1}{m_1}+\frac{\tilde{\tau}_2}{m_2}\right)$. Below, we discuss the range of temperatures in which the ambipolar effect gives only a sub-leading correction to the thermal conductivity for the Hubbard-like and screened Coulomb interactions.

Recently,  it was shown in Ref.~\cite{Zarenia20} that a non-uniform charge distribution can form near the contacts used for thermal conductivity measurements. As a consequence, the ambipolar contribution to thermal conductivity should not diverge even in the clean limit. The intriguing result of Ref.~\cite{Zarenia20} suggests that in the presence of disorder, the ambipolar effect can be suppressed up to even higher temperatures than naively expected. 

Keeping only the leading terms, and using $\langle\!\langle\xi^2_{\bf p}\rangle\!\rangle=\pi^2 T^2/3$, one obtains the low-temperature result
\begin{align}
\kappa=\frac{\pi^2}{3} \mathcal{N} T\left(\frac{\tilde{\tau}_1}{m_1}+\frac{\tilde{\tau}_2}{m_2}\right).\label{eq:kappaKeyes}
\end{align}
and the Lorenz ratio
\begin{align}
\mathcal{L}&=\frac{\pi^2\tilde{m}}{3e^2}\left(\frac{1}{\tau}+\frac{1}{\tau_{el}}\right)\left(\frac{\tilde{\tau}_1}{m_1}+\frac{\tilde{\tau}_2}{m_2}\right).
\label{eq:LKeyes}
\end{align}
As is obvious from Eq.~\eqref{eq:LKeyes}, the Wiedemann-Franz law prediction $\mathcal{L}=\mathcal{L}_0=\pi^2/(3e^2)$ can be expected to hold only for vanishing intraband and interband scattering rates, i.e. for $T\rightarrow 0$.

\section{Fermi-liquid collision integral}
\label{sec:FLCI}

In this section, we use the eigenfunction expansion method first introduced in Refs.~\cite{Brooker68, Jensen68} and employed for disordered systems in Refs.~\cite{Bennett69,Lee20} to find solutions of the linearized Boltzmann equations \eqref{BoltzmannEq_Linearized} with intraband and interband collision integrals of the Fermi-liquid type. We obtain expressions for the electric and thermal conductivies of a compensated metal which will form the basis for further discussions in Sec.~\ref{sec:discussion}. 

\subsection{Electron-electron collision integral}

We write the collision integral as
\begin{align}
I_{ee}^{cc'}\{f\} 
& = - \frac{f_{c,{\bf p}}}{\tau_{out,{\bf p}}^{cc'}} + \frac{1 - f_{c,{\bf p}}}{\tau_{in,{\bf p}}^{cc'}},
\label{Coll_Int_ee_1}
\end{align}
where out- and in-scattering rates take the form
\begin{align}
\frac{1}{\tau_{out,{\bf p}}^{cc'}}
& = \int_{{\bf q},{\bf p'},{\bf q'}} W_{{\bf p}{\bf q},{\bf p}'{\bf q}'}^{cc'} \delta(\epsilon_{c,{\bf p}} + \epsilon_{c',{\bf q}} - \epsilon_{c,{\bf p}'} - \epsilon_{c',{\bf q}'}) 
\nonumber\\
& ~~~ \times \delta({\bf p} + {\bf q} - {\bf p}' - {\bf q}') f_{c',{\bf q}} (1 - f_{c,{\bf p}'}) (1 - f_{c',{\bf q}'}),
\label{OutscatteringRate}
\\
\frac{1}{\tau_{in,{\bf p}}^{cc'}}
& = \int_{{\bf q},{\bf p'},{\bf q'}} W_{{\bf p}'{\bf q}',{\bf p}{\bf q}}^{cc'} \delta(\epsilon_{c,{\bf p}} + \epsilon_{c',{\bf q}} - \epsilon_{c,{\bf p}'} - \epsilon_{c',{\bf q}'})
\nonumber\\
& ~~~ \times \delta({\bf p} + {\bf q} - {\bf p}' - {\bf q}') f_{c,{\bf p}'} f_{c',{\bf q}'} (1 - f_{c',{\bf q}}).
\end{align}
Here, $W$ describes the probability for electrons with momentum ${\bf p}$ in band $c$ and ${\bf q}$ in $c'$ to be scattered into states with ${\bf p}'$ in $c$ and ${\bf q}'$ in $c'$,
\begin{align}
W_{{\bf p}{\bf q},{\bf p}'{\bf q}'}^{cc'} 
& = s (2\pi)^{d+1} |U_{{\bf p}{\bf q},{\bf p}'{\bf q}'}^{cc'}|^2.
\label{eq:W}
\end{align}
In this formula, $s$ counts spin degeneracy, and $U_{{\bf p}{\bf q},{\bf p}'{\bf q}'}^{cc'}$ is the matrix element describing the interaction between carriers in bands $c$ and $c'$. Spin-triplet scattering processes have been ignored for the sake of simplicity. If needed, they can be included into the formalism straightforwardly \cite{Sykes70,Li18}.

Following a standard procedure \cite{Ziman01}, we linearize the collision integral with the ansatz $\delta f_{c,{\bf p}}=\beta n_{F}(\xi_{c,{\bf p}}) [1 - n_{F}(\xi_{c,{\bf p}})] \Phi_{c,{\bf p}}$, to obtain the form
\begin{align}
I_{ee}^{cc'}\{\Phi\}
& = - \frac{1}{T} \int_{{\bf q},{\bf p}',{\bf q}'} W_{{\bf p}{\bf q},{\bf p}'{\bf q}'}^{cc'} \delta(\epsilon_{c,{\bf p}} + \epsilon_{c',{\bf q}} - \epsilon_{c,{\bf p}'} - \epsilon_{c',{\bf q}'}) 
\nonumber\\
& ~~~ \times \delta({\bf p} + {\bf q} - {\bf p}' - {\bf q}') n_{F}(\xi_{c,{\bf p}}) n_{F}(\xi_{c',{\bf q}})
\nonumber\\
& ~~~ \times [1 - n_{F}(\xi_{c,{\bf p}'})] [1 - n_{F}(\xi_{c',{\bf q}'})] 
\nonumber\\
& ~~~ \times (\Phi_{c,{\bf p}} + \Phi_{c',{\bf q}} - \Phi_{c,{\bf p}'} - \Phi_{c',{\bf q}'}), 
\label{Coll_Int_ee_2}
\end{align}
and further make the dependence on the external fields explicit by writing $\Phi_{c,{\bf p}}$ as $\Phi_{c,{\bf p}} = \sum_{\alpha\in\{E,T\}} \phi_c^\alpha(\xi_{c,{\bf p}}) \boldsymbol{v}_{c,{\bf p}} \cdot  {\bf F}_\alpha$ with ${\bf F}_E = - e {\bf E}$ and ${\bf F}_T = -\nabla_{\bf r} T$ with four unknown functions $\phi_1^E(\xi_{1,{\bf p}})$, $\phi_2^E(\xi_{2,{\bf p}})$, $\phi_1^T(\xi_{1,{\bf p}})$, $\phi_2^T(\xi_{2,{\bf p}})$. 

From now on, we focus on a three-dimensional compensated metal with Fermi-liquid interactions, at temperatures in the degenerate regime $T\ll\epsilon_{F}$. 

\subsection{Exact solution}
\label{subsec:exact}
The general collision integral \eqref{Coll_Int_ee_2} can be simplified considerably in the degenerate regime, as we describe in Appendix \ref{Appendix:FLCollisionIntegral}. The resulting approximate form of the electron-electron collision integral reads as~\cite{Remark1}
\begin{align}
& I_{ee}^{cc}\{\phi\}
= - \frac{m_c^3}{(2\pi)^7T} n_{F}(\xi_{c,{\bf p}})[1-n_{F}(\xi_{c,{\bf p}})] 
\nonumber\\
& ~ \times \int_{-\infty}^\infty d\omega K(\omega,\xi_{c,{\bf p}}) \sum_{\alpha\in \{E,T\}} \boldsymbol{v}_{c,{\bf p}} \cdot {\bf F}_{\alpha} 
\nonumber\\
& ~ \times \big[ \langle\tilde{W}_{cc}(\theta,\varphi)\rangle_{\rm av} \big\{ \phi^{\alpha}_{c,s}(\xi_{c,{\bf p}}) + \phi^{\alpha}_{c,a}(\xi_{c,{\bf p}}) - \phi^{\alpha}_{c,s}(\xi_{c,{\bf p}} + \omega) \big\} 
\nonumber\\
& ~ - \langle \tilde{W}_{cc}(\theta,\varphi) (1+2\cos\theta) \rangle_{\rm av} \phi^{\alpha}_{c,a}(\xi_{c,{\bf p}}+\omega) \big],
\label{Coll_Int_ee_diag}
\end{align}
\begin{align}
& I_{ee}^{c\bar{c}}\{\phi\}
= - \frac{m_c m_{\bar{c}}^2}{(2\pi)^7T} n_{F}(\xi_{c,{\bf p}})[1-n_{F}(\xi_{c,{\bf p}})] 
\nonumber\\
& ~ \times \int_{-\infty}^\infty d\omega K(\omega,\xi_{c,{\bf p}}) \sum_{\alpha\in \{E,T\}} \boldsymbol{v}_{c,{\bf p}} \cdot {\bf F}_{\alpha} 
\nonumber\\
& ~ \times \big[ \langle\tilde{W}_{c\bar{c}}(\theta,\varphi)\rangle_{\rm av} \big\{ \phi^{\alpha}_{c,s}(\xi_{c,{\bf p}}) + \phi^{\alpha}_{c,a}(\xi_{c,{\bf p}}) \big\} 
\nonumber\\
& ~ - \langle\tilde{W}_{c\bar{c}}(\theta,\varphi) \cos\Theta \rangle_{\rm av} \big\{ \phi^{\alpha}_{c,s}(\xi_{c,{\bf p}} + \omega) + \phi^{\alpha}_{c,a}(\xi_{c,{\bf p}} + \omega) \big\} 
\nonumber\\
& ~ + \langle \tilde{W}_{c\bar{c}}(\theta,\varphi) (1 - \cos\Theta) \rangle_{\rm av} \frac{m_c}{m_{\bar{c}}} \phi^{\alpha}_{\bar{c},s}(\xi_{c,{\bf p}}+\omega) 
\nonumber\\
& ~ + \langle \tilde{W}_{c\bar{c}}(\theta,\varphi) (1 + 2\cos\theta - \cos\Theta) \rangle_{\rm av} \frac{m_c}{m_{\bar{c}}} \phi^{\alpha}_{\bar{c},a}(\xi_{c,{\bf p}}+\omega) \big],
\label{Coll_Int_ee_offdiag}
\end{align}
Here, it was convenient to split $\phi^\alpha$ into symmetric and antisymmetric parts $\phi^{\alpha}(\xi)=\phi^{\alpha}_s(\xi)+\phi^{\alpha}_a(\xi)$, where $\phi^{\alpha}_s(\xi)=\phi^{\alpha}_s(-\xi)$ and $\phi^{\alpha}_a(\xi)=-\phi^{\alpha}_a(-\xi)$. The function $K(\omega,\xi_{c,{\bf p}})$ contains information on the occupation of states involved in the scattering processes,
\begin{align}
K(\omega,\xi_{c,{\bf p}}) 
= \omega n_{B}(\omega)\frac{1-n_{F}(\xi_{c,{\bf p}}+\omega)}{1-n_{F}(\xi_{c,{\bf p}})},
\end{align}
where $n_{B}(\omega) = [\exp(\beta\omega) - 1]^{-1}$ is the Bose-Einstein distribution. 
For later reference, we note the relations $K(-\omega,-\xi_{c,{\bf p}}) = K(\omega,\xi_{c,{\bf p}})$, and $\int d\omega K(\omega, \xi_{c,{\bf p}}) = [\xi_{c,{\bf p}}^2+(\pi T)^2]/2$. The scattering probability $\tilde{W}$ is obtained from $W$ by fixing all incoming and outgoing momenta to $p_F$, and is characterized by two angles: $\theta$ is the angle between the two incoming momenta, and $\varphi$ is the angle between the two planes spanned by the incoming momenta and by the outgoing momenta. The third angle $\Theta$ is defined by the relation $\cos\Theta = 1 - 2\sin^2(\theta/2) \cos^2(\varphi/2)$. The angular average is defined by $\langle X(\theta,\varphi) \rangle_{\rm av} = \int d(\Omega/4\pi) X(\theta,\varphi) / \cos(\theta/2)$ with the differential solid angle $d\Omega = \sin\theta d\theta d\varphi$.

The sum of the intraband and interband collision integrals, Eqs.~\eqref{Coll_Int_ee_diag} and \eqref{Coll_Int_ee_offdiag}, can be written in the compact form
\begin{align}
I_{ee,c}\{\phi\}
& = - \frac{4n_{F}(\xi_{c,{\bf p}})[1-n_{F}(\xi_{c,{\bf p}})]}{\pi^2 T^3 \tau_{out,c}} \int_{-\infty}^\infty d\omega K(\omega,\xi_{c,{\bf p}}) 
\nonumber\\
& ~~~ \times \sum_{\alpha\in \{E,T\}} \boldsymbol{v}_{c,{\bf p}} \cdot {\bf F}_{\alpha} \sum_{\gamma\in\{s,a\}}
\big[ \phi^\alpha_{c,\gamma}(\xi_{c,{\bf p}}) 
\nonumber\\
& ~~~ - \Lambda_{c,\gamma} \phi^\alpha_{c,\gamma}(\xi_{c,{\bf p}} + \omega) + \Xi_{c,\gamma} \phi^\alpha_{\bar{c},\gamma}(\xi_{c,{\bf p}}+\omega) \big].
\label{Coll_Int_ee_tot}
\end{align}
Here, $1/\tau_{out,c}$ denotes the out-scattering rate defined in Eq.~\eqref{OutscatteringRate}, evaluated on the Fermi surface and in equilibrium,
\begin{align}
\frac{1}{\tau_{out,c}}
& = (u_{cc} + u_{c\bar{c}}) \frac{T^2}{\epsilon_{F}},
\label{OutScatteringRate}
\end{align}
with the dimensionless parameters 
\begin{align}
u_{cc} & = \frac{m_c^3 \epsilon_{F}}{2^9\pi^5} \langle\tilde{W}_{cc}(\theta,\varphi)\rangle_{\rm av},
\\
u_{c\bar{c}} & = \frac{m_c m_{\bar{c}}^2 \epsilon_{F}}{2^9\pi^5} \langle\tilde{W}_{c\bar{c}}(\theta,\varphi)\rangle_{\rm av}.
\end{align}
Additional dimensionless parameters $\Lambda_{c,s/a}$ and $\Xi_{c,s/a}$, relevant to the symmetric and antisymmetric parts of $\phi_c^\alpha$ in Eq.~\eqref{Coll_Int_ee_tot}, respectively, are defined by
\begin{align}
\Lambda_{c,s}
& = 1 - \frac{m_{\bar{c}}}{m_c} \Xi_{c,s}
\label{Lambda_cs}
\\
\Xi_{c,s}
& = \frac{(m_{\bar{c}}/m_c) \langle \tilde{W}_{c\bar{c}}(\theta,\varphi) (1 - \cos\Theta) \rangle_{\rm av}}{\langle\tilde{W}_{cc}(\theta,\varphi)\rangle_{\rm av} + (m_{\bar{c}}^2/m_c^2) \langle\tilde{W}_{c\bar{c}}(\theta,\varphi)\rangle_{\rm av}},
\label{Xi_cs}
\\
\Lambda_{c,a}
& = \frac{\langle\tilde{W}_{cc}(\theta,\varphi) (1 + 2\cos\theta) \rangle_{\rm av}}{\langle\tilde{W}_{cc}(\theta,\varphi)\rangle_{\rm av}} - \frac{m_{\bar{c}}}{m_c} \Xi_{c,a},
\label{Lambda_ca}
\\
\Xi_{c,a}
& = \frac{(m_{\bar{c}}/m_c) \langle \tilde{W}_{c\bar{c}}(\theta,\varphi) (1 + 2\cos\theta - \cos\Theta) \rangle_{\rm av}}{\langle\tilde{W}_{cc}(\theta,\varphi)\rangle_{\rm av} + (m_{\bar{c}}^2/m_c^2) \langle\tilde{W}_{c\bar{c}}(\theta,\varphi)\rangle_{\rm av}}.
\label{Xi_ca}
\end{align}

For a {\it clean} compensated metal, the two coupled Boltzmann equations \eqref{BoltzmannEq_Linearized} with electron-electron collision integral given by Eq.~\eqref{Coll_Int_ee_tot} were solved in Ref.~\cite{Li18} using the eigenfunction expansion originally introduced for Fermi liquids in Refs.~\cite{Jensen68,Brooker68}. The approach of Refs.~\cite{Jensen68,Brooker68} was also used to study conventional disordered metals in Refs.~\cite{Bennett69,Lee20}. Here, we derive solutions of \eqref{BoltzmannEq_Linearized} for disordered compensated metals with the electron-electron collision integral given in Eq.~\eqref{Coll_Int_ee_tot}. We find the electric and thermal conductivities by inserting the solutions into Eq.~\eqref{CurrentDensity}. The technical details of this procedure are described in Appendix \ref{Appendix:SolvingBoltzmannEquation}. Here, we summarize the main results. The expressions for the electric and thermal conductivities read 
\begin{align}
\sigma
& = \sum_{c\in\{1,2\}} \sum_{n=0}^\infty \tilde{\sigma}_c \Upsilon_{c,n}^\sigma (2n + \varepsilon_c + 1/2) 
\nonumber\\
& ~~~ \times \frac{\lambda_{\bar{c},2n} - \Lambda_{\bar{c},s} - \Xi_{c,s} \tau_{out,\bar{c}}/\tau_{out,c}}{(\lambda_{c,2n} - \Lambda_{c,s})(\lambda_{\bar{c},2n} - \Lambda_{\bar{c},s}) - \Xi_{c,s} \Xi_{\bar{c},s}},
\label{ElectricCond_General}
\\
\kappa
& = \sum_{c\in\{1,2\}} \sum_{n=0}^\infty \tilde{\kappa}_c \Upsilon_{c,n}^\kappa (2n + \varepsilon_c + 3/2)
\nonumber\\
& ~~~ \times \frac{\lambda_{\bar{c},2n+1} - \Lambda_{\bar{c},a} - \Xi_{c,a} \tau_{out,\bar{c}}/\tau_{out,c}}{(\lambda_{c,2n+1} - \Lambda_{c,a})(\lambda_{\bar{c},2n+1} - \Lambda_{\bar{c},a}) - \Xi_{c,a} \Xi_{\bar{c},a}},
\label{ThermalCond_General}
\end{align}
respectively. Here, we defined $\tilde{\sigma}_c = \mathcal{N}e^2\tau_{ei,c}/m_c$ and $\tilde{\kappa}_c = \pi^2\mathcal{N}T\tau_{ei,c}/(3m_c)$, the contributions of band $c$ to the electric and thermal conductivities in the absence of electron-electron scattering, respectively. We further introduced the functions $\lambda_{c,n} = (n+\varepsilon_c) (n+\varepsilon_c+1)/2$, and
\begin{align}
\Upsilon_{c,n}^\sigma 
& = \frac{\tau_{out,c}}{8\tau_{ei,c}} \frac{\Gamma(n+1/2)\Gamma(n+\varepsilon_c+1/2)[\Gamma(n+(\varepsilon_c+1)/2)]^2}{\Gamma(n+1)\Gamma(n+\varepsilon_c+1)[\Gamma(n+\varepsilon_c/2+1)]^2},
\\
\Upsilon_{c,n}^\kappa 
& = \frac{3\tau_{out,c}}{8\tau_{ei,c}} \frac{\Gamma(n+3/2)\Gamma(n+\varepsilon_c+3/2)[\Gamma(n+(\varepsilon_c+1)/2)]^2}{\Gamma(n+1)\Gamma(n+\varepsilon_c+1)[\Gamma(n+\varepsilon_c/2+2)]^2}.
\end{align}
These results will be discussed in the next section.

\subsubsection{Intraband scattering dominant regime}

An instructive special case is obtained in the intraband scattering dominant regime. For $\tilde{W}_{cc} \gg \tilde{W}_{c\bar{c}}$, parameters simplify considerably: $1/\tau_{out,c} = u_{cc} T^2/\epsilon_{\rm F}$, $\Xi_{c,s} = \Xi_{c,a} = 0$, $\Lambda_{c,s} = 1$, and
\begin{align}
\Lambda_{c,a}
& = \frac{\langle\tilde{W}_{cc}(\theta,\varphi) (1 + 2\cos\theta) \rangle_{\rm av}}{\langle\tilde{W}_{cc}(\theta,\varphi)\rangle_{\rm av}}.
\end{align}
The parameter $\Lambda_{c,a}$ takes values between $-1$ and $3$. Head-on collisions $(\theta=\pi)$ correspond to $\Lambda_{c,a}=-1$, and collinear scattering ($\theta=0$) to $\Lambda_{c,a}=3$.

In the intraband scattering dominant regime, transport occurs independently in electron and hole bands. The electric conductivity is simply described by the Drude formula~\cite{Remark2}
\begin{align}
\sigma
& = \tilde{\sigma}_1 + \tilde{\sigma}_2.
\label{ElectricCond_Intra}
\end{align}
The electric conductivity is insensitive to intraband electron-electron collisions when interband scattering processes are negligible, it depends only on the electron-impurity scattering rate. This is because intraband collisions do not relax the total momentum, and therefore do not affect the electric current directly. Intraband electron-electron collisions can relax the thermal current, however, and this is why the thermal conductivity takes a nontrivial form
\begin{align}
\kappa
& = \sum_{c\in\{1,2\}} \sum_{n=0}^\infty \tilde{\kappa}_c \Upsilon_{c,n}^\kappa \frac{2n + \varepsilon_c + 3/2}{\lambda_{c,2n+1} - \Lambda_{c,a}}.
\label{ThermalCond_Intra}
\end{align}
A detailed discussion of Eq.~\eqref{ThermalCond_Intra} and a comparison with the simple results obtained from the RTA can be found in Ref.~\cite{Lee20} for the closely related single-band case. Let us note here that due to the intraband scattering the thermal conductivity is always reduced with respect to to the Drude result $\kappa=\tilde{\kappa}_1+\tilde{\kappa}_2$, and as a consequence $\mathcal{L}/\mathcal{L}_0<1$.

\subsubsection{Interband scattering dominant regime}
\label{subsec:inter}

We now turn to the interband scattering dominant regime. For $\tilde{W}_{cc} \ll \tilde{W}_{c\bar{c}}$ as well as (intraband) electron-impurity scatterings suppressed, the electric and thermal conductivities are greatly simplified into
\begin{align}
\frac{\sigma}{\tilde{\sigma}_0}
& = \sum_{n=0}^\infty \frac{4n + 3}{4(n+1)(2n+1)[(n+1)(2n+1) + 1 - 2\tilde{\Lambda}]},
\label{ElectricCond_Inter}
\\
\frac{\kappa}{\tilde{\kappa}_0}
& = \sum_{n=0}^\infty \frac{3(4n + 5)}{4(n+1)(2n+3)[(n+1)(2n+3) + \tilde{\Xi} - 2\tilde{\Lambda}]}.
\label{ThermalCond_Inter}
\end{align}
For the purpose of normalization, we defined the conductivities
\begin{align}
\tilde{\sigma}_0
& = \frac{2^9\pi^5 \mathcal{N}e^2}{m_1^2 m_2^2 T^2 \langle\tilde{W}_{12}(\theta,\varphi)\rangle_{\rm av}},
\\
\tilde{\kappa}_0
& = \frac{2^9\pi^7 \mathcal{N}}{3 m_1^2 m_2^2 T \langle\tilde{W}_{12}(\theta,\varphi)\rangle_{\rm av}},
\end{align}
with $\tilde{\kappa}_0/(\tilde{\sigma}_0 T)=\mathcal{L}_0$. We further introduced the dimensionless parameters
\begin{align}
\tilde{\Lambda}
& = \frac{\langle\tilde{W}_{12}(\theta,\varphi) \cos\Theta \rangle_{\rm av}}{\langle\tilde{W}_{12}(\theta,\varphi)\rangle_{\rm av}},
\\
\tilde{\Xi}
& = \frac{\langle \tilde{W}_{12}(\theta,\varphi) (1 + 2\cos\theta) \rangle_{\rm av}}{\langle\tilde{W}_{12}(\theta,\varphi)\rangle_{\rm av}}.
\end{align}

The temperature dependence of the electric and thermal conductivities is fully determined by that of $\tilde{\sigma}_0$ and $\tilde{\kappa}_0$, the sums in Eqs.~\eqref{ElectricCond_Inter} and \eqref{ThermalCond_Inter} provide prefactors which depend crucially on the angular dependence of the scattering probabilities. The normalized Lorenz ratio $\mathcal{L}/\mathcal{L}_0$ is solely determined by the ratio of the two sums, and is therefore temperature-independent and in general different from 1.

\section{Discussion}
\label{sec:discussion}

In the previous section we derived exact expressions for the electric and thermal conductivities of compensated metals in the presence of disorder and Fermi-liquid-type interactions, Eqs.~\eqref{ElectricCond_General} and \eqref{ThermalCond_General}. Moreover, we obtained simplified results for two limiting cases: the case where intraband scattering dominates over interband collisions and the opposite limit. We turn now to a discussion of these results for systems where electron-electron collisions are due to: (i) Hubbard-like interactions or (ii) screened Coulomb interactions.

The exact formulas for the conductivities, Eqs.~\eqref{ElectricCond_General} and \eqref{ThermalCond_General}, requires knowledge of the angular dependence of the interaction, which is not easily accessible in experimental systems. By contrast,  the conductivities obtained by the RTA method [given by Eqs.~\eqref{Sigma_EE},~\eqref{eq:kappaKeyes} and~\eqref{eq:LKeyes}] are fully specified by a few phenomenological parameters. Thus, establishing the connection between the two approaches may provide a simple way to analyze and understand experimental findings. It is important to note that the Fermi-liquid integrals only retain contributions to the leading order in temperature, $T^2/\epsilon_{\rm F}^2$. Consequently, for the comparison we omit the subleading ambipolar contribution to the thermal conductivity and Lorenz ratio in the RTA approach.

\begin{figure}[t]
\centering
\includegraphics[width=0.45\textwidth]
{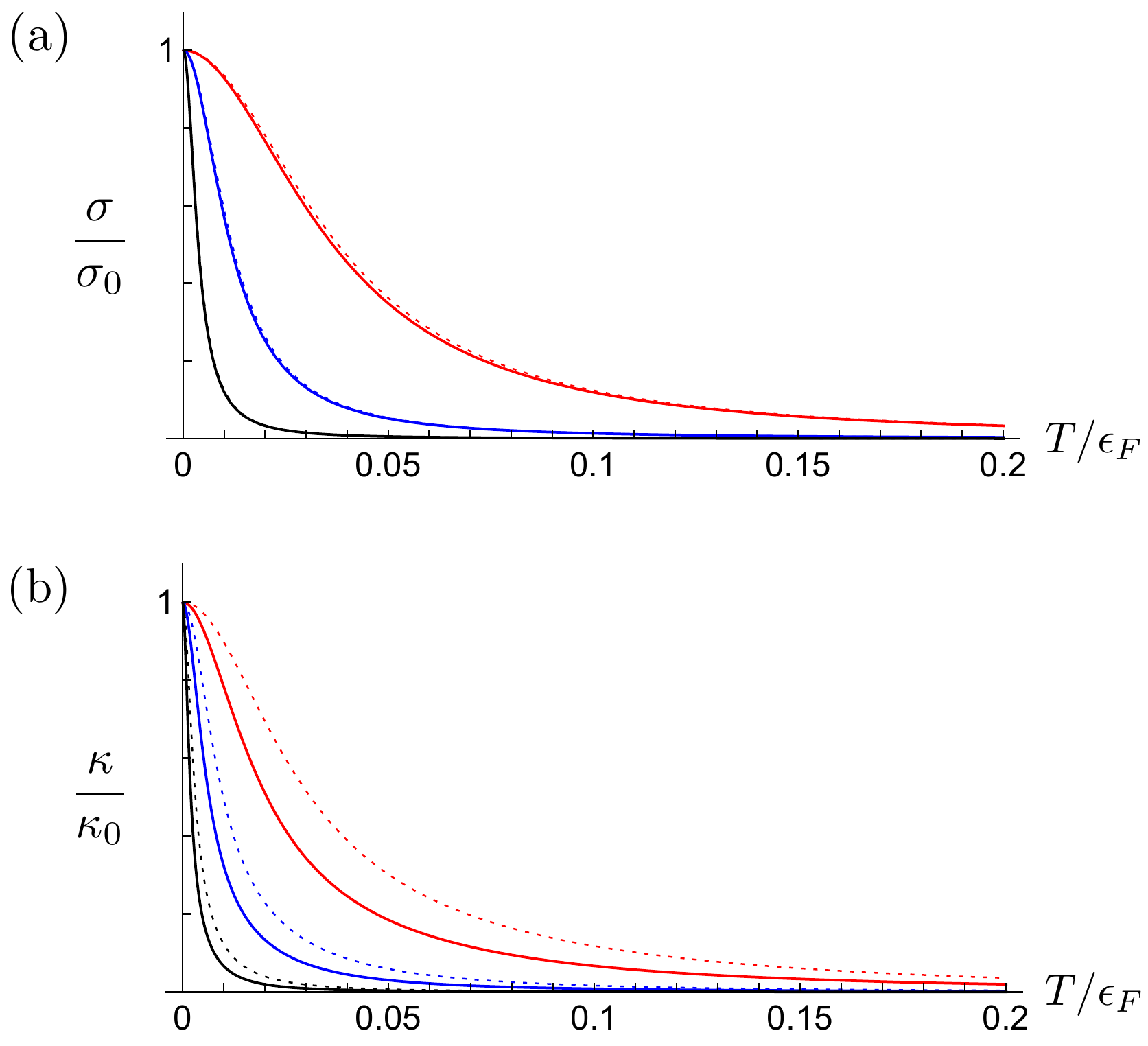} \\
\caption{
Transport coefficients in the presence of Hubbard-like interactions as a function of temperature for different strengths of impurity scattering:  The electric conductivity in units of $\sigma_0 = \mathcal{N}e^2 (\tau_{ei,1}/m_1 + \tau_{ei,2}/m_2)$ is shown in (a), and  the thermal conductivity in units of $\kappa_0 = (\pi^2 \mathcal{N}T/3) (\tau_{ei,1}/m_1 + \tau_{ei,2}/m_2)$ is plotted in (b). We used the following values for the calculation of the exact expressions for the transport coefficients:  $u_{11} = 1.29$, $\tilde{W}_{11} = \tilde{W}_{22}=1$, $\tilde{W}_{12} / \tilde{W}_{11} = \tilde{W}_{21} / \tilde{W}_{22} = 2$, $m_2/m_1=1.5$, and $\tau_{ei,2}/\tau_{ei,1} = 10$. The red, blue, and black curves correspond to $1/ \epsilon_{\rm F}\tau_{ei,1}$ = 0.1,  0.01, 0.001, respectively. The dotted lines indicate the solutions to the phenomenological RTA approach,  Eqs.~\eqref{Sigma_EE},~\eqref{eq:kappaKeyes} and~\eqref{eq:LKeyes}, by matching the relaxation times as described in the main text. The poor fit to $\kappa$ illustrates the different effect that interband scattering has on the electric and thermal transport (see discussion in the main text).}  
\label{Figure2a}
\end{figure}

\begin{figure}[t]
\centering
\includegraphics[width=0.45\textwidth]
{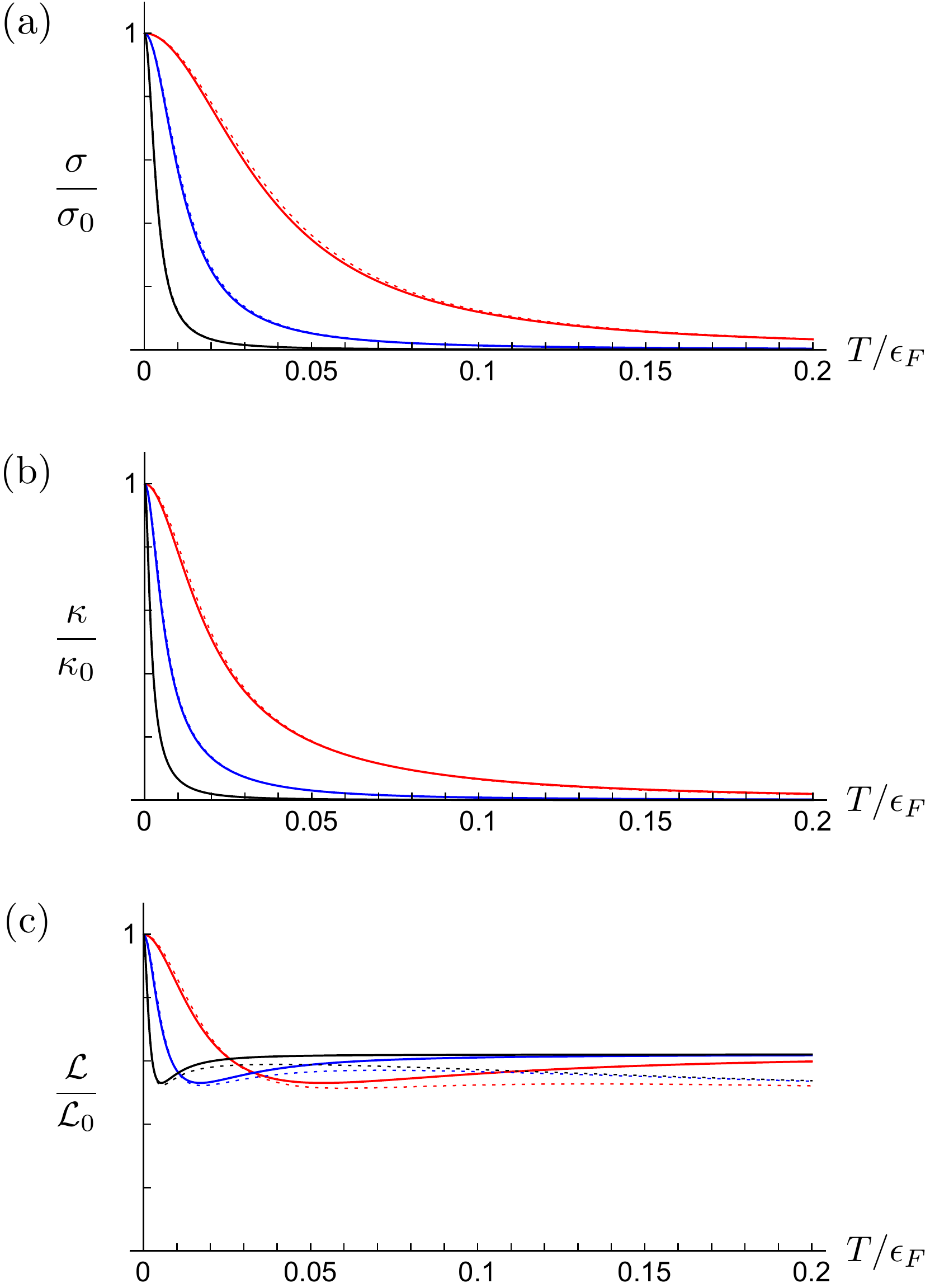} \\
\caption{Transport coefficients in the presence of Hubbard-like interactions as a function of temperature for different strengths of impurity scattering: The parameters used for the plots are identical to those in Fig.~\ref{Figure2a}. Here, however, the dotted lines indicate the solutions to the phenomenological RTA approach with {\it two} independent interband-scattering times as given by Eqs.~\eqref{eq:sigmaGRTA} and \eqref{eq:kappaGRTA}. The ambipolar contribution has been dropped for comparison. The Lorenz ratio $\mathcal{L}(T)$ in units of $\mathcal{L}_0 = \pi^2/(3e^2)$ is also displayed.}    
\label{Figure2}
\end{figure}

We first analyze the case of Hubbard-like short-range interactions that scatter uniformly over all Fermi-surface angles. In Fig.~\ref{Figure2a}, we plot the electric and thermal conductivities found using the exact solution and the RTA for a compensated metal in the presence of Hubbard-like interactions. An important step in comparing the solutions is to determine the scattering times that enter the RTA as phenomenological parameters. Here, we find the relaxation time for electron-impurity scattering, $\tau_{el}$, by matching the results for the electric conductivity at $T=0$, where interaction effects go to zero. In the absence of disorder, the decay time of the electric current, $\tau_E$, is determined by interband scattering events alone. Consequently, we fix this time by comparing the RTA expression for the electric conductivity~\eqref{Sigma_EE} to the exact result obtained for the clean interband-scattering-dominated regime in Sec.~\eqref{subsec:inter}. Finally, to find the intraband relaxation time, we similarly fit the two solutions for $\kappa$ in the clean limit and in the absence of interband collisions. As illustrated in Fig.~\ref{Figure2a}, the two solutions for the electric conductivity agree very well, but this is not the case for the thermal conductivity. Alternatively, we could determine both the intra- and interband relaxation times using the expressions for the thermal conductivity. Then, the RTA would only poorly fit the exact solution for the electric conductivity.

The failure of the phenomenological solutions to match both $\sigma$ and $\kappa$ using one set of relaxation times occurs because the RTA does not take into account that electron-electron interactions affect the two conductivities differently. For example, forward scattering events are more effective in suppressing the heat flow than relaxing the charge current. This shortcoming can be easily mitigated by introducing different inelastic scattering times into the RTA expressions for the electric and thermal conductivities
 \begin{align}
\sigma&=\frac{\mathcal{N}e^2}{\tilde{m}}\left(\frac{1}{\tau_{el}}+\frac{1}{\tau_{E}}\right)^{-1}
\label{eq:sigmaGRTA},\\
\kappa&=\frac{\pi^2}{3}\mathcal{N}T \left(\frac{\tilde{\tau}_1}{m_1}+\frac{\tilde{\tau}_2}{m_2}\right)
\label{eq:kappaGRTA},
\end{align}
with $\tau_{el}$ defined in Eq.~\eqref{eq:tauel} and
\begin{align}
\tilde{\tau}_c=\left(\frac{1}{\tau_{ei,c}}+\frac{1}{\tau_{T,ee}^{cc}}+\frac{\tilde{m}}{m_c\tau_{T}}\right)^{-1}
\label{eq:taucGTRA}.
\end{align}
As a result, the phenomenological expression for the Lorenz ratio becomes
\begin{align}
\frac{\mathcal{L}}{\mathcal{L}_0}=\left(\frac{1}{\tau_{el}}+\frac{1}{\tau_E}\right)\tilde{m}\left(\frac{\tilde{\tau}_1}{m_1}+\frac{\tilde{\tau}_2}{m_2}\right).
\label{eq:LorenzGRTA}
\end{align}
In Appendix~\ref{app:ansatz} we provide a transparent example for the necessity to distinguish $\tau_E$ and $\tau_T$, the analysis of the forward-scattering dominated regime with the ansatz of constant $\phi^E$ and $\phi^T/\xi$. The solutions obtained from this ansatz take precisely the form of Eqs.~\eqref{eq:sigmaGRTA} and \eqref{eq:kappaGRTA}. 

As illustrated in Fig.~\ref{Figure2}, the modified RTA expressions given above agree well with the exact solution and capture details such as the minimum of the Lorenz ratio (and its position). The phenomenological parameters $\tau_{el} $, $\tau_E$ and $\tau_{T,ee}^{cc}$ are determined as prescribed above for  the case of a single interband relaxation time. To set the value of $\tau_T$, we match  between Eq.~\eqref{eq:kappaGRTA} and the exact solution found in Sec.~\eqref{subsec:inter} in the clean limit and in the absence of intraband scattering.  We give a detailed description of this procedure in Appendix~\ref{Appendix:BC}. Interestingly, in the solutions obtained from the RTA the elastic and inelastic scattering rates are independent; they enter the conductivity additively, unlike in the exact solution, Eqs.~\eqref{ElectricCond_General} and \eqref{ThermalCond_General}. Nevertheless, \eqref{eq:sigmaGRTA} and \eqref{eq:kappaGRTA} may often provide a satisfactory approximation for any form of the interaction. Equipped with the modified RTA solutions, we are now ready to address the  temperature dependence of the Lorenz ratio in the presence of Hubbard-like and screened Coulomb interactions.

\subsection{Hubbard interaction}


The exact solutions for the conductivities  include angular averages over the scattering amplitudes. Thus, they considerably simplify for the Hubbard-like interactions. Specifically,  Eqs.~\eqref{Lambda_cs} and \eqref{Xi_cs} become
\begin{align}
\Lambda_{c,s}
& = \frac{3 + (m_{\bar{c}}^2/m_c^2) (\tilde{W}_{c\bar{c}}/\tilde{W}_{cc})}{3[1 + (m_{\bar{c}}^2/m_c^2) (\tilde{W}_{c\bar{c}}/\tilde{W}_{cc})]},
\label{Lambda_Hubbard}
\\
\Xi_{c,s}
& = \frac{2 (m_{\bar{c}}/m_c) (\tilde{W}_{c\bar{c}}/\tilde{W}_{cc})}{3[1 + (m_{\bar{c}}^2/m_c^2) (\tilde{W}_{c\bar{c}}/\tilde{W}_{cc})]}.
\label{Xi_Hubbard}
\end{align}
In addition, Eqs.~\eqref{Lambda_ca} and \eqref{Xi_ca} result in $\Lambda_{c,a} = 1/3$ and $\Xi_{c,a} = 0$. We use the above expression for the calculation of the transport coefficients. In the clean limit, the Lorenz ratio is independent of temperature and it satisfies the following properties: (i) $\mathcal{L}$ becomes independent of $m_2/m_1$ (ii) It approaches zero when the strength of intraband interactions  significantly exceeds the interband one, $\tilde{W}_{cc}\gg\tilde{W}_{c\bar{c}}$. (iii) In the opposite limit, $\tilde{W}_{cc}\ll\tilde{W}_{c\bar{c}}$, the Lorenz number approaches a maximum value of $\mathcal{L}/\mathcal{L}_0 \approx 0.930$ (iv) In the maximally symmetric case $\tilde{W}_{11} = \tilde{W}_{22} = \tilde{W}_{12} = \tilde{W}_{21}$, we obtain $\mathcal{L}/\mathcal{L}_0 \approx 0.489$~\cite{Remark3}.

In Fig.~\ref{Figure2} we show the electric conductivity, the thermal conductivity, and the Lorenz ratio of a compensated metal with Hubbard interactions for  several values of  the disorder scattering strength. As $T\rightarrow 0$ the inelastic scattering rates become small, and both the electric and thermal conductivities attain their Drude values, i.e., $\sigma_0 = \mathcal{N}e^2 (\tau_{ei,1}/m_1 + \tau_{ei,2}/m_2)$ and $\kappa_0 = (\pi^2 \mathcal{N}T/3) (\tau_{ei,1}/m_1 + \tau_{ei,2}/m_2)$. Hence, $\mathcal{L}(T=0)=\mathcal{L}_0$ in agreement with the Wiedemann-Franz law~\cite{Wiedemann1853}. As the temperature increases, inelastic collisions become important, and both conductivities decrease. The range of temperatures over which the conductivities exhibit the largest drop is set by the ratios between the elastic and inelastic collision rates and it gets wider for larger electron-impurity scattering. The drop in the thermal conductivity at $T>0$ is stronger than for  its electric counterpart. The difference between the temperature dependence of $\sigma(T)$ and $\kappa(T)$ is due to the fact that interband collisions are more effective at relaxing the heat current than the electric current, $\tau_E/\tau_T\approx 2.15$  (see Appendix \ref{Appendix:BC}), while intraband collisions hardly affect the charge flow. Consequently, at non-zero temperature, $\mathcal{L}(T)$ becomes smaller than $\mathcal{L}_0$. The Lorenz ratio saturates to its clean-limit value~\cite{Li18} at temperatures where  inelastic scattering is significantly stronger than electron-impurity collisions.  In Fig.~\ref{Figure2}, we see that the decrease in $\mathcal{L}(T)$ is non-monotonic and all curves have a minimum. While the Lorenz ratio at its minimum is independent of the disorder strength, its position moves to higher temperatures for larger $1/\tau_{el}$.

To understand the origin of the minimum in $\mathcal{L}(T)$, it is instructive to examine the expressions for the conductivities found within the RTA method, Eqs.~\eqref{eq:sigmaGRTA} and~\eqref{eq:kappaGRTA}.  Both conductivities can be written as a sum of the contributions from each band. For the electric conductivity, however, both contributions depend only on global relaxation times, $\tau_{el}$ and $\tau_{E}$. As a result, the slope of the decreasing  $\sigma(T)$ is set by a single parameter, the ratio between the two relaxation times,  $\tau_{el}/\tau_{E}$.  By contrast, the contribution to the thermal conductivity from each band cannot be written in terms of the global relaxation times, and the slope of $\kappa(T)$ depends on two independent parameters. A minimum of the Lorenz ratio appears if $\kappa'(T)/\kappa(T)$ changes from being larger than $\sigma'(T)/\sigma(T)$ at low $T$ to being smaller at high $T$. Such a scenario can occur in systems with  unequal masses and/or impurity scattering rates for the two bands.  This is the case in all three curves presented in Fig.~\ref{Figure2}.

\begin{figure}[tb]
\centering
\includegraphics[width=0.45\textwidth]
{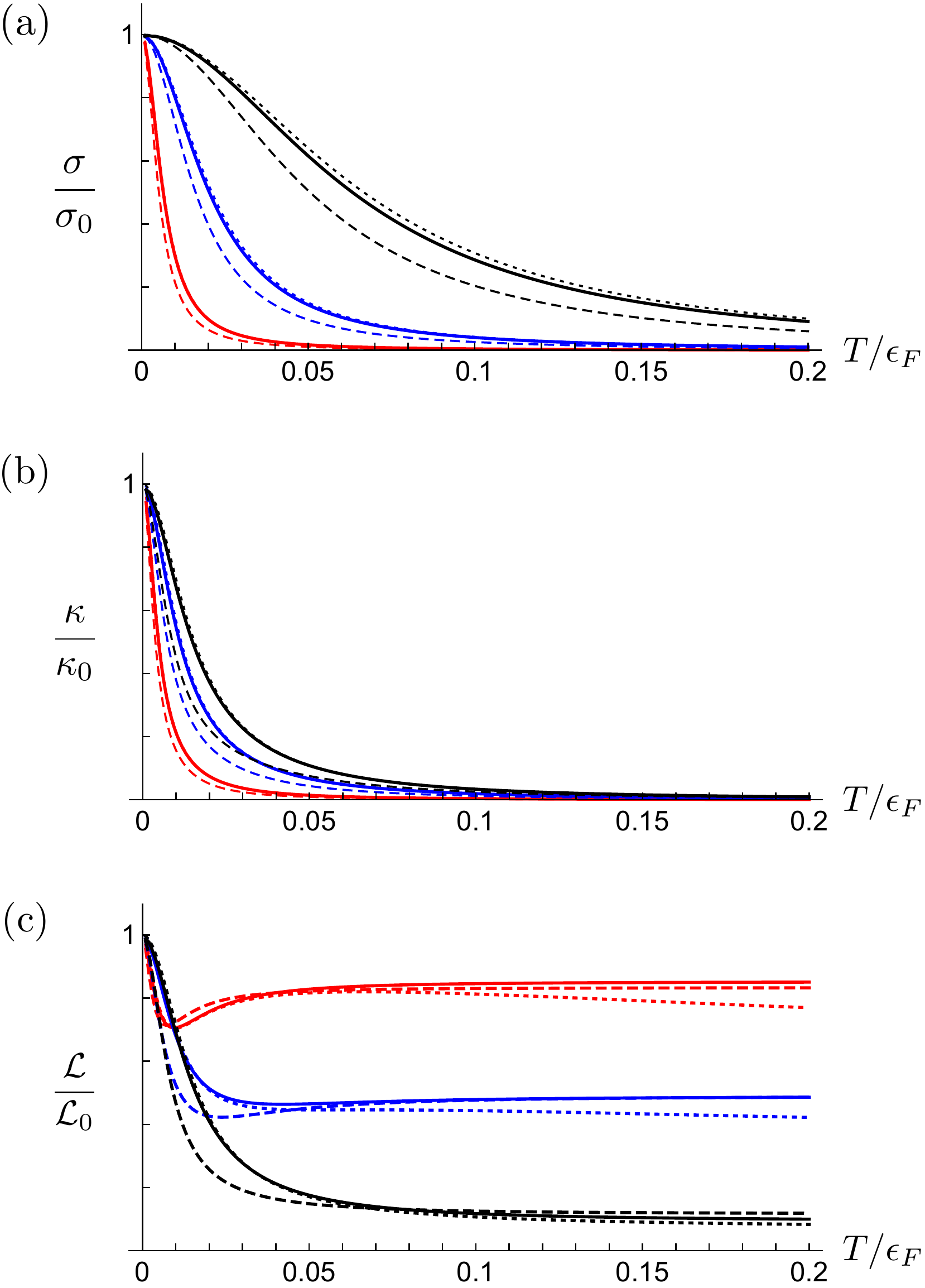} \\
\caption{
Transport coefficients for Hubbard-type interactions as a function of temperature for different strength of interband scattering. The parameters used for the exact solutions are: $\gamma_{ei,1} / \epsilon_{\rm F} = 0.01$, $\gamma_{ei,2} / \epsilon_{\rm F} = 0.001$, $u_{11} = 1.29$, and $\tilde{W}_{11} = \tilde{W}_{22}$. The black, blue and red  curves correspond to $\tilde{W}_{12} / \tilde{W}_{11} = \tilde{W}_{21} / \tilde{W}_{22} = 0.1,~1,~10$, respectively. We consider bands of equal masses (solid line) and bands with different masses $m_2/m_1=1.5$ (dashed lines). The dotted lines illustrate the RTA expressions for the transport coefficient (with $m_1=m_2$).}
\label{Figure1}
\end{figure}

To highlight the different role of intraband and interband scattering on the Lorenz ratio, we plot the conductivities and $\mathcal{L}(T)$ for different values of  $\tilde{W}_{c\bar{c}}$ in Fig.~\ref{Figure1}.  We obtain that for the parameters chosen here, increasing the strength of interband interactions has a much stronger effect on the temperature dependence of $\sigma(T)$ than on $\kappa(T)$. This is expected as long as intraband collisions are important, since then interband scattering determines the thermal conductivity in combination with intraband scattering, while interband scattering alone determines the electric conductivity.  Figure~\ref{Figure1} (c) demonstrates that a minimum in the Lorenz ratio is not universal within the Fermi-liquid approximation.

%
    
\begin{figure}[t]
\centering
\includegraphics[width=0.45\textwidth]
{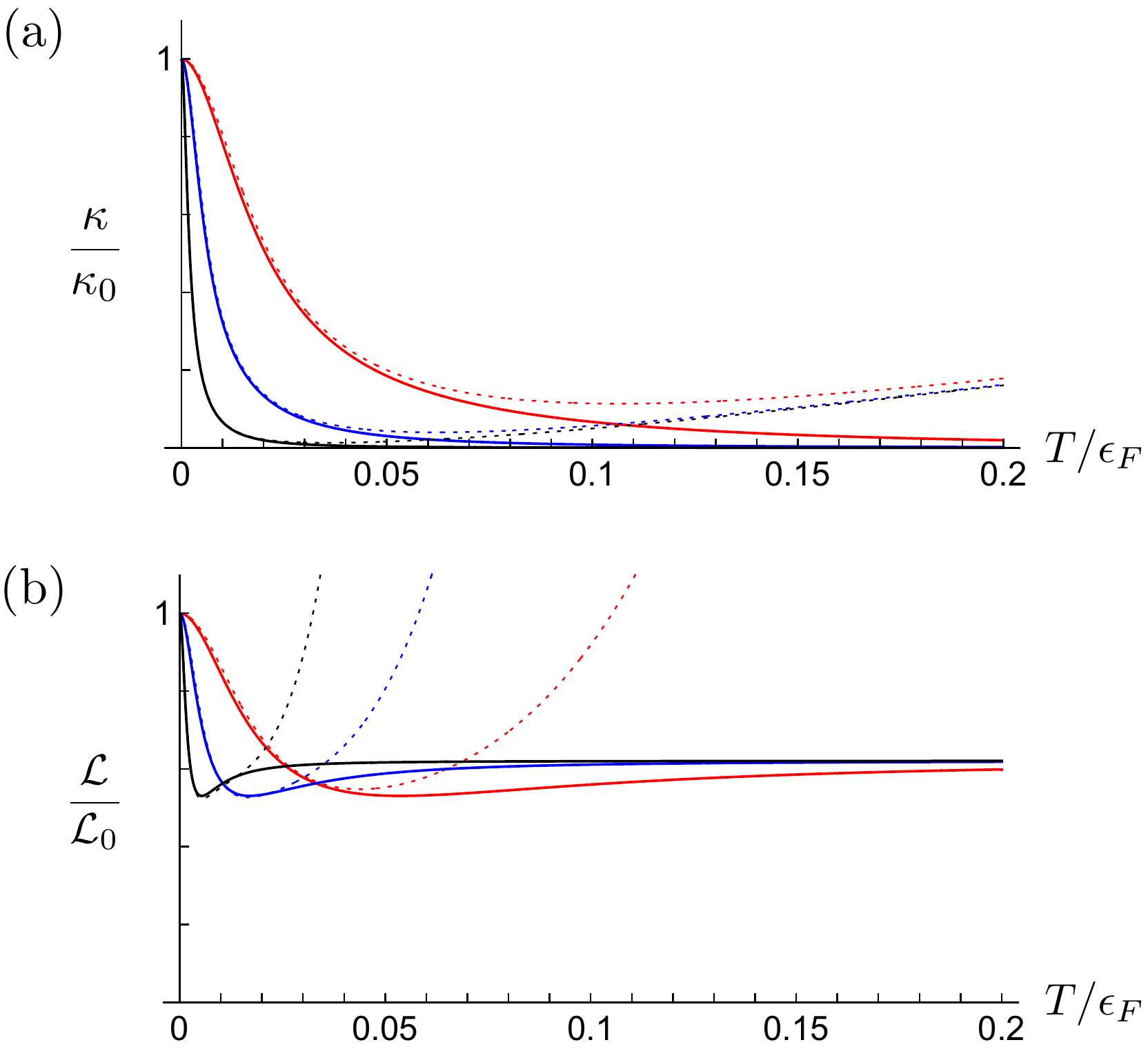} \\
\caption{Transport coefficients in the presence of Hubbard-like interactions as a function of temperature for different strengths of impurity scattering. Parameters are chosen as in Fig.~\ref{Figure2}. In contrast to Fig.~\ref{Figure2}, the results of the phenomenological RTA approach (dotted lines), include the ambipolar contribution.}  
\label{Figure3A}
\end{figure}

The ambipolar contribution that was found within the RTA solution for the thermal conductivity diverges as $T\rightarrow\infty$. For $T/\epsilon_F\ll1$, this contribution is of the order of $\kappa_0T^2/\epsilon_F^2$ [see Eq.~\eqref{ThermalConductivity}]. Thus, the ambipolar contribution is negligible as long as $\kappa/\kappa_0$  is larger than $T^2/\epsilon_F^2$. For the parameters used in Fig.~\ref{Figure2}, the ambipolar contribution becomes relevant only for temperatures exceeding the minimum in the Lorenz ratio (see also Fig.~\ref{Figure3A}). In general, the ambipolar contribution, which is unique for compensated metals, suggests that a minimum in the Lorenz number should always be present, even in the absence of phonons. 
  
\subsection{Screened Coulomb interaction}

\begin{figure}[t]
\centering
\includegraphics[width=0.45\textwidth]
{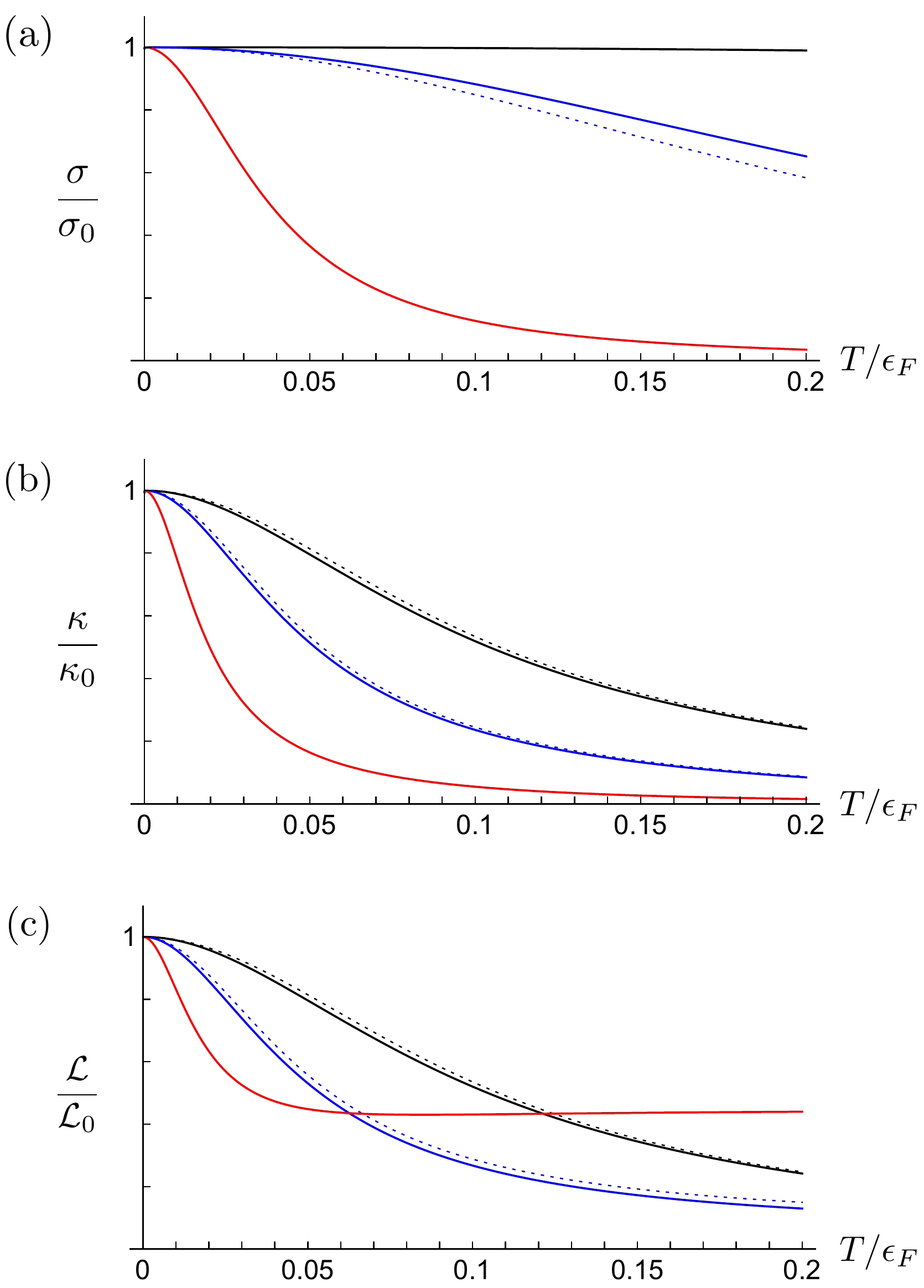} \\
\caption{
Transport coefficients for the screened Coulomb interaction as a function of temperature for different screening wave numbers. Here, we  set $\gamma_{ei,1} / \epsilon_{F} = 0.01$, $\gamma_{ei,2} / \epsilon_{F} = 0.001$ and $m_1=m_2$. The black, blue, red colored curves correspond to $k_{TF}/p_{F} = 0.1,~0.4,~4$, respectively. The dotted lines are the RTA expressions that have been obtained  by matching scattering times in the clean forward scattering limit as explained in the main text.
}
\label{Figure4}
\end{figure}

\begin{figure}[t]
\centering
\includegraphics[width=0.45\textwidth]
{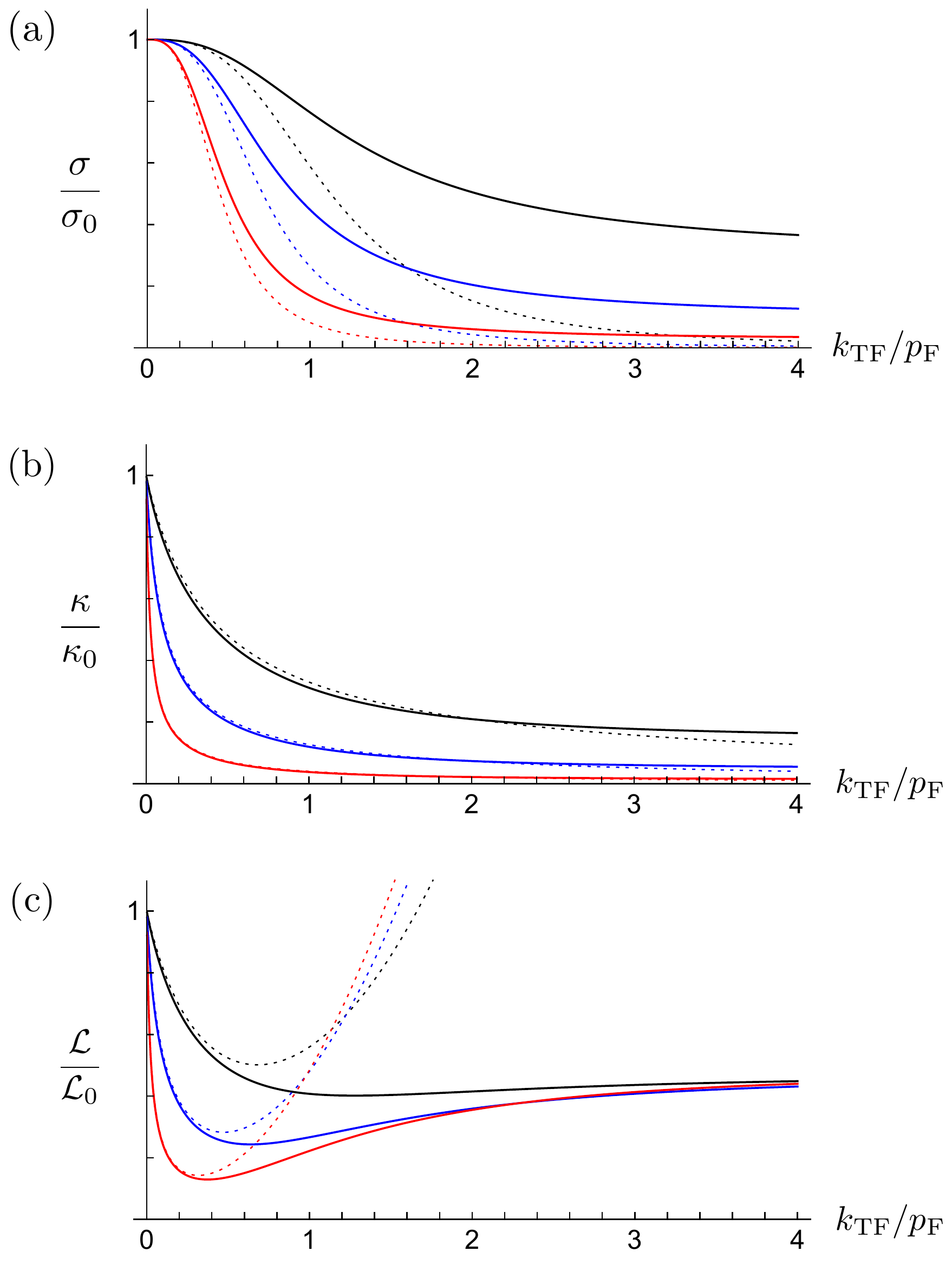} \\
\caption{
Transport coefficients for the screened Coulomb interaction as a function of $k_{{TF}}$  for different temperatures. We use here the same parameters as in Fig.~\ref{Figure4}. In all panels, black, blue and red curves correspond to $T/\epsilon_{F} = 0.05,~0.1,~0.2$, respectively. The solid lines indicate the exact solution, while the RTA results are given by the dotted lines. We found the phenomenological parameters entering the RTA calculation by taking the limit of small  $k_{TF}$. Thus, the dotted curves start to strongly deviate from the solid lines as $k_{TF}/p_{F}\rightarrow1$.
  }
\label{Figure5}
\end{figure}

The discussion above focused on short-range interactions that give rise to uniform scattering over the Fermi surface. We now examine the opposite limit of strongly angle-dependent scattering. Specifically, we study  the transport coefficients in the presence of  screened Coulomb interactions 
\begin{align}
U(k) = \frac{1}{\epsilon_0}\frac{4\pi e^2}{k^2 + k_{TF}^2},
\label{CoulombInteraction}
\end{align}
where the Thomas-Fermi (TF) screening wave number $k_{TF}$ is given by $k_{TF}^2 = 4\pi e^2 [\nu_1(\epsilon_{\rm F}) + \nu_2(\epsilon_{\rm F})] /\epsilon_0= 4e^2 (m_1+m_2) p_{F}/(\pi\epsilon_0)$, and $\epsilon_0$ is the dielectric constant. The electron-electron collisions are dominated by forward scattering in the limit of   weak screening, $k_{TF}\ll p_{\rm F}$, and they become almost momentum/angle-independent in the opposite limit $k_{TF}\gg p_F$. Strictly speaking the  interaction given by Eq.~\eqref{CoulombInteraction} is valid only for $k_{TF}<p_F$. Nevertheless, we use it for arbitrary $k_{TF}$ as a model interaction to study the effect of changing the angle dependence of the inelastic collisions on the transport coefficients and the Lorenz ratio.

To evaluate the electric and thermal conductivities, Eqs.~\eqref{ElectricCond_General} and \eqref{ThermalCond_General}, we insert Eq.~\eqref{CoulombInteraction} into Eqs.~\eqref{OutScatteringRate}-\eqref{Xi_ca}.  Parameterizing the momentum as $k = 2p_{\rm F} \sin(\theta/2)\cos(\varphi/2)$ and averaging over the angle yields \begin{align}
\frac{1}{\tau_{out,c}}
& = \frac{\pi^3}{32} \frac{m_c}{m_1} \frac{1 + m_2^2/m_1^2}{(1 + m_2/m_1)^2} \frac{T^2}{\epsilon_{\rm F}}
\nonumber\\
& ~~~ \times \tilde{k}_{TF} \bigg[ \tan^{-1}\bigg(\frac{1}{\tilde{k}_{TF}}\bigg) + \frac{\tilde{k}_{TF}}{1 + \tilde{k}_{TF}^2} \bigg],
\end{align}
\begin{align}
\Xi_{c,s}
& = \frac{2m_{\bar{c}}/m_c}{1 + m_{\bar{c}}^2/m_c^2} \frac{(1 + \tilde{k}_{TF}^2) \tan^{-1}(1/\tilde{k}_{TF}) - \tilde{k}_{TF}}{(1 + \tilde{k}_{TF}^2) \tan^{-1}(1/\tilde{k}_{TF}) + \tilde{k}_{TF}} \tilde{k}_{TF}^2,
\\
\Lambda_{c,a}
& = \frac{ (1 - 2\tilde{k}_{TF}^2) (1 + \tilde{k}_{TF}^2) \tan^{-1}(1/\tilde{k}_{TF}) + \tilde{k}_{TF} (1 + 2\tilde{k}_{TF}^2)}{(1 + \tilde{k}_{TF}^2) \tan^{-1}(1/\tilde{k}_{TF}) + \tilde{k}_{TF}},
\end{align}
and $\Xi_{c,a} = 0$. Here, we defined the dimensionless parameter $\tilde{k}_{TF} = k_{TF}/(2p_{\rm F})$. 

An instructive limiting case arises in the clean limit of the forward scattering regime, $\tilde{k}_{TF}\ll 1$. There,  the leading contributions to the thermal and electric conductivities become (see Appendix~\ref{Appendix:BC} and Ref.~\cite{Li18})
\begin{align}
\sigma&=\frac{6}{\pi^4}\frac{\mathcal{N}e^2}{T^2}\frac{(m_1+m_2)^2}{m_1^2m_2^2}\frac{p_F^2}{\tilde{k}_{TF}^3},\label{eq:sforward}\\
\kappa&=\frac{4}{\pi^2}\frac{\mathcal{N}}{T}\frac{(m_1+m_2)^2}{m_1^2m_2^2}\frac{p_F^2}{\tilde{k}_{TF}},\label{eq:kappaforward}
\end{align}
so that $\mathcal{L}/\mathcal{L}_0=k_{TF}^2/2 p_F^2$ is temperature-independent. 

To compare the exact solution for the transport coefficients with the RTA conductivities, Eqs.~\eqref{eq:sigmaGRTA} and \eqref{eq:kappaGRTA}, we apply the matching procedure discussed above to extract the phenomenological scattering times. We find that $1/\tau_E\propto \tilde{k}_{TF}^3$, $1/\tau_T\propto \tilde{k}_{TF}$ and  $1/\tau_{T,ee}^{cc}\propto \tilde{k}_{TF}$ in the limit of weak screening $\tilde{k}_{TF}\ll1$ (see Appendix \ref{Appendix:BC}). Figure~\ref{Figure4} illustrates the temperature dependence of the conductivities and the Lorenz ratio for different values of $\tilde{k}_{TF}$. For very weak screening (black and blue curves with $k_{TF}/p_F=0.1$ and $k_{TF}/p_F=0.4$, respectively), the inelastic interactions barely relax the electric current and $\sigma$ is almost independent of temperature. By contrast, $\kappa$ is significantly suppressed with increasing temperature. The difference between the temperature dependence of the two transport coefficients is also reflected by the ratio between the two relaxation times, $\tau_E/\tau_T=1/(2\tilde{k}_{TF}^2)$, which can become large for small $\tilde{k}_{TF}$.  As a result of the weak temperature dependence of $\sigma(T)$, the Lorenz ratio acquires its temperature dependence solely from $\kappa$,  and does not exhibit a minimum. For strongly screened interactions  (red curves with  $\tilde{k}_{TF}=4$), the conductivities resemble those found in the previous subsection for the Hubbard-like interactions. Without electron-impurity scattering, the Lorenz ratio becomes smaller with increasing screening length. Interestingly, this no longer holds in the presence of disorder. As illustrated in  Fig.~\ref{Figure4} (c), $\mathcal{L}(T)$ at low temperatures is suppressed with increasing $k_{TF}$. When electron-electron collisions dominate over  impurity scattering, the curves cross, and as expected for the clean limit, the Lorenz ratio decreases with  increasing $k_{TF}$. Finally, we note that the RTA approximation works well for low $k_{TF}/p_F$, as expected. We do not include here the RTA result for $k_{TF}/p_F=4$ because it is well outside the range of validity  for the scattering times found in Appendix~\ref{Appendix:BC}.

In Fig.~\ref{Figure5}, we show the dependence of the  transport coefficients on the inverse screening length, $k_{TF}$, at fixed temperature.  As can be seen from Fig.~\ref{Figure5} (c), for each temperature the minimal Lorenz ratio is obtained at different values of  $k_{TF}/p_F$. This minimum becomes more shallow with growing temperature. As expected, the RTA solutions (dashed lines) match the curves at low $\tilde{k}_{TF}$, and strongly deviate from the exact result as  $\tilde{k}_{TF}$ approaches unity.

\begin{figure}
\centering
\includegraphics[width=0.45\textwidth]{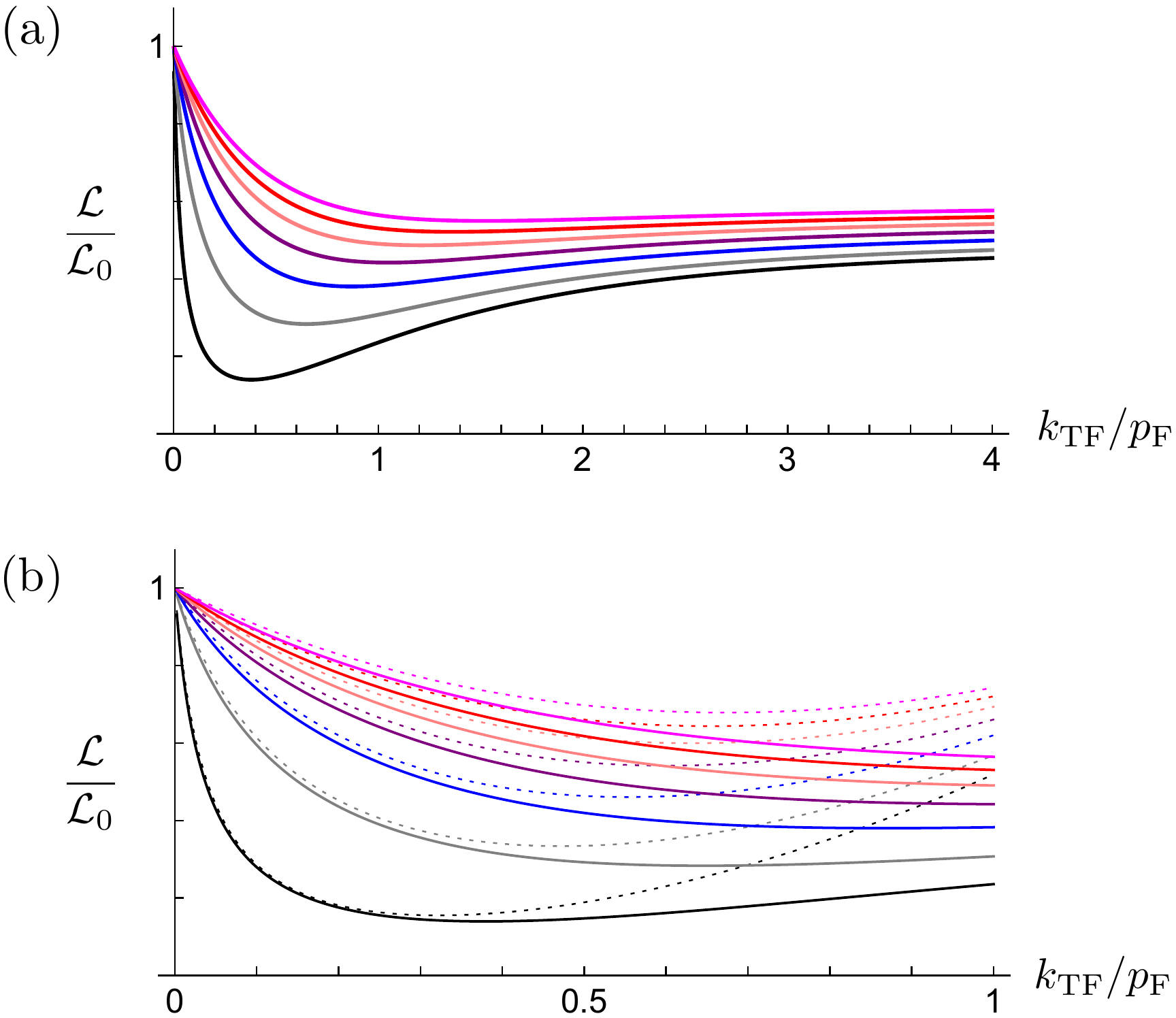}\\
\caption{
Lorenz ratio in the presence of  screened Coulomb interaction as a function of $k_{TF}$ for varying disorder scattering strength. For all curves  $m_1=m_2$, $T/\epsilon_F=0.05$, and $\gamma_{ei,1}=\gamma_{ei,2}$. The black, gray, blue, purple, pink, red, and magenta lines correspond to $\gamma_{ei,c}/\epsilon_F=\{1,4,8,12,16,20,24\}\times 10^{-4}$, respectively. The dotted lines represent the RTA solutions in the limit of small $\tilde{k}_{TF}$.  
}
\label{Figure6}
\end{figure}

\begin{figure}
\centering
\includegraphics[width=0.45\textwidth]{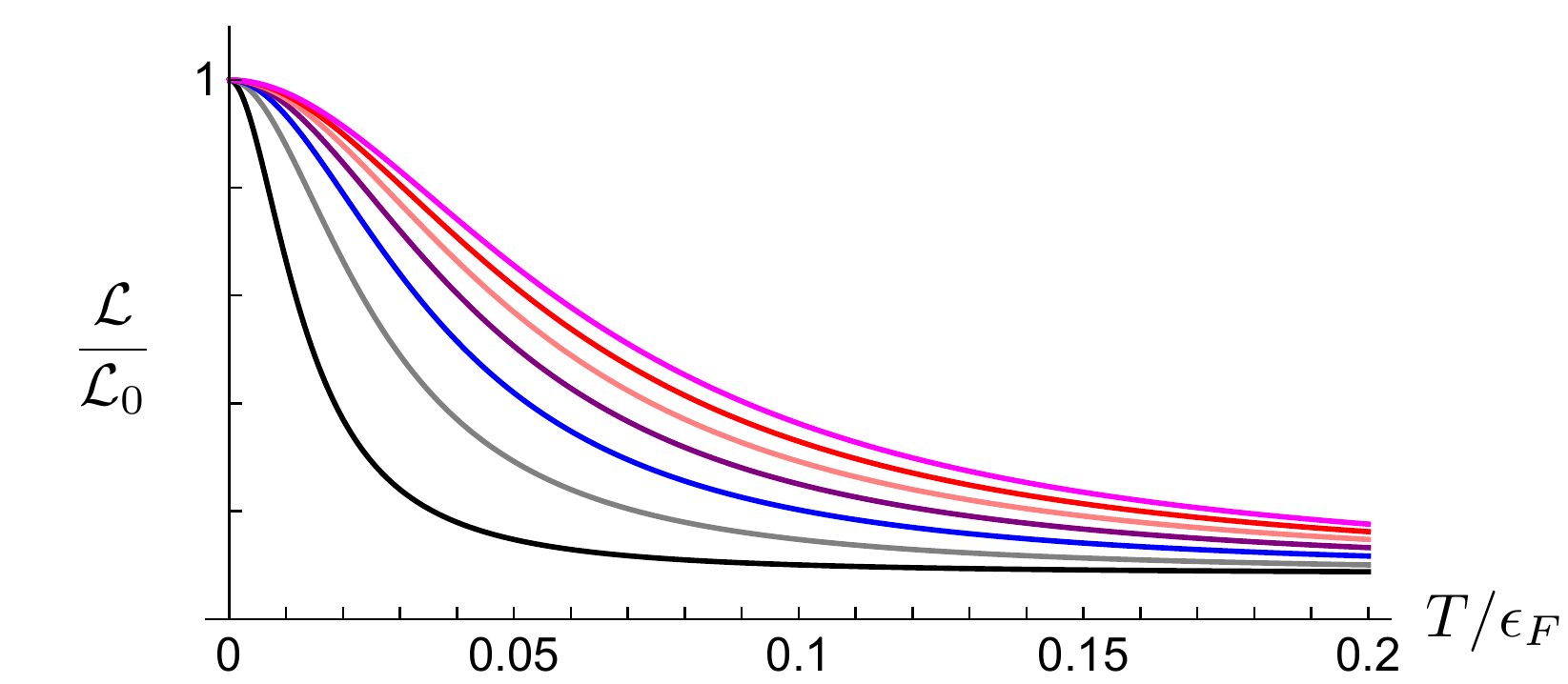}\\
\caption{
Lorenz ratio in the presence of screened Coulomb interaction as a function of temperature for varying disorder scattering strength. Here, we set $m_1=m_2$, $k_{TF}/p_{\rm F}=0.5$, and $\gamma_{ei,1}=\gamma_{ei,2}$. The black, gray, blue, purple, pink, red, and magenta curves correspond to $\gamma_{ei,c}/\epsilon_F=(1,4,8,12,16,20,24)\times 10^{-4}$, respectively. 
}
\label{Figure7}
\end{figure}

For completeness, we present in Figs.~\ref{Figure6} and ~\ref{Figure7} the Lorenz ratio for different levels of disorder as functions of  $k_{TF}/p_F$ and temperature, respectively. Disorder significantly enhances the Lorenz ratio at low $T$, but becomes irrelevant at  high $T$.  For small $\tilde{k}_{TF}$, we find a good agreement  between the RTA (dashed lines) and the exact (solid lines) solutions, see Fig.~\ref{Figure6} (b). Therefore, we can use the RTA expression to estimate the screening length for which the Lorenz ratio is minimal. This is especially simple when the bands are symmetric, since the RTA expression becomes
\begin{align}
\frac{\mathcal{L}}{\mathcal{L}_0}=\frac{\frac{1}{\tau}+\frac{\pi^4}{24}\frac{T^2}{\epsilon_F} \tilde{k}^3_{TF}}{\frac{1}{\tau}+\frac{\pi^4}{48}\frac{T^2}{\epsilon_F}\tilde{k}_{TF}}.
\label{eq:Lorenzsimple}
\end{align}
At small $\tilde{k}_{TF}$, the denominator grows faster than the numerator, while at larger $\tilde{k}$ these roles reverse. In between, the minimum in $\mathcal{L}$ occurs.

\section{Conclusion}
\label{sec:conclusion}

We studied the effect of both disorder and electron-electron interactions on electric and thermal transport in compensated metals by solving the Boltzmann kinetic equation.  The eigenfunction expansion method was used to solve the Boltzmann equation within the Fermi-liquid approximation and to find exact expressions for the conductivities.  In the limit $T\rightarrow 0$, electron-electron collision terms vanish and the electric and thermal conductivities satisfy the Wiedemann-Franz law.  As temperature is increased, inelastic scattering becomes important and, as long as phonons can be neglected, both conductivities decrease. The electric conductivity is mainly affected by interband collisions, the Baber mechanism, while both intra- and interband collisions contribute to the suppression of the thermal conductivity with increasing temperature. As a result,  the drop in $\kappa(T)$ is, in general, stronger than for $\sigma(T)$, and the Lorenz ratio also reduces with increasing temperature. Within the Fermi-liquid approximation and as long as phonon contributions can be neglected, both conductivities are expected to go to zero as $1/T^2$ at high temperatures where the electron-electron scattering rates significantly exceed those of the inelastic collisions.  The Lorenz ratio, by contrast, saturates to a non-zero value, $0<\mathcal{L}<\mathcal{L}_0$.  To gain more insight, we also analyzed the kinetic equation in the RTA, and found closed form solutions for the transport coefficients in terms of the phenomenological scattering rates, Eqs.~\eqref{eq:sigmaGRTA} and \eqref{eq:kappaGRTA}. We developed a scheme for calculating the scattering rates from the exact expressions for the conductivities. We showed that the transport coefficients calculated using the two methods agree over a large range of temperatures. The transparency and simplicity of the phenomenological approach and its connection to the microscopic models, may prove valuable for the interpretation of future transport experiments in compensated metals.

Our main result, the exact formulas for the conductivities given by Eqs.~\eqref{ElectricCond_General} and \eqref{ThermalCond_General}, are expressed in terms of averages of the inelastic collision probabilities over angles on the Fermi surface. To illustrate the implications of the angular averaging, we examined two types of interactions: (i) the short-range Hubbard-like and (ii) the long-range screened Coulomb. In particular, we studied the dependence of the Lorenz ratio on the screening length. In agreement with the analysis of Refs.~\cite{Li18, Jaoui18}, we obtained that without disorder $\mathcal{L}$ becomes smaller as the screening gets weaker. This is no longer valid in the presence of impurities. As a result of both elastic and inelastic collisions,  the Lorenz ratio becomes a non-monotonic function of the inverse screening length with a minimum at finite $k_{\rm TF}$.  In systems with asymmetric bands, $\mathcal{L}(T)$ can be also non-monotonic in temperature, and go through a minimum at intermediate temperatures even in the absence of phonons. This minimum is, however, not universal and it is highly sensitive to the properties of the bands as well as to the form and the strength of the interactions. 

The temperature dependence of our solution for the Lorenz ratio is (qualitatively) consistent with the experimental observations of Refs.~\cite{Jaoui18,Gooth18}. We derived the transport coefficients only to the leading order in $T/\epsilon_F\ll1$, i.e., in the low-temperature limit. Specifically, the leading contributions to the thermal conductivity and the Lorenz ratio do not include the ambipolar effect, which  in the presence of disorder is $O(\kappa_0T^2/\epsilon_F^2)$. To the leading order,  $\kappa(T)/\kappa_0$ equals unity at $T=0$ and reduces to a smaller value at higher temperatures, while the subleading ambipolar contribution grows with increasing $T$. Thus, the latter can be neglected as long as   $\kappa(T)/\gg \kappa_0T^2/\epsilon_F^2$. This condition is valid near the minimum of the Lorenz ratio found~\cite{Jaoui18}  in WP$_2$ near $T_m\sim 10$ K.
(The Fermi energies of the electron and hole pockets in this material are on the order of $1000$ K \cite{Schoenemann17}.)  It was recently shown~\cite{Zarenia20} that the ambipolar effect can be small even in clean systems as a result of a non-uniform charge distribution accumulating near the system boundaries.  In the presence of disorder,  a similar mechanism can further reduce the already small ambipolar contribution. Scattering by impurities was essential in our calculation to capture the reduction of the Lorenz ratio with temperature, as seen in experiments. In addition to the ambipolar effect, we also neglected corrections to the Lorenz ratio due to phonons. These were found~\cite{Jaoui18} to be small at temperatures $\lesssim T_m$. At higher temperatures, electron-phonon scattering and the ambipolar contribution to the thermal conductivity can become significant~\cite{RemarkPhonons}.

\acknowledgments

We thank A.~M.~Finkel'stein for discussions and useful comments. This work was supported by the College of Arts and Sciences at the University of Alabama (W.~L., G.~S.) and the National Science Foundation (NSF) under Grant No. DMR-1742752 (G.~S.) and the Army Research Office (ARO) under Grant No. W911NF2010013 and W911NF-16-1-0182 (W.~L.).  K.~M.~was supported by Grant No. 2017608 from the United States-Israel Binational Science Foundation (BSF).

\appendix

\begin{widetext}


\section{Derivation of the approximate form of the electron-electron collision integral, Eq.~\eqref{Coll_Int_ee_offdiag} }
\label{Appendix:FLCollisionIntegral}

In this Appendix, we derive the approximate form of the electron-electron collision integral presented in Eq.~\eqref{Coll_Int_ee_diag} and \eqref{Coll_Int_ee_offdiag} in the main text. The case $c=c'$ is a rather straightforward generalization of the single-band case discussed in Ref.~\cite{Lee20}, here we focus our attention on $c\ne c'$. We study the case of a Fermi liquid in three dimensions ($d=3$)~\cite{Abrikosov59}, adjusted to the presence of two bands in a compensated metal. Our starting point is Eq.~\eqref{Coll_Int_ee_2}. In a first step, we introduce the energy $\omega$ transferred during electron-electron scattering with the help of the identity $\delta(\epsilon_{c,{\bf p}} + \epsilon_{c',{\bf q}} - \epsilon_{c,{\bf p}'} - \epsilon_{c',{\bf q}'}) = \int_{-\infty}^\infty d\omega \delta(\omega - \epsilon_{c,{\bf p}'} + \epsilon_{c,{\bf p}}) \delta(\omega - \epsilon_{c',{\bf q}} + \epsilon_{c',{\bf q}'})$. This identity allows us to present the collision integral in the following form
\begin{align}
I_{ee}^{cc'}\{\phi\} 
& = - \frac{1}{T} \sum_{\alpha\in \{E,T\}} \int_{{\bf q},{\bf p'},{\bf q'}} \int_{-\infty}^\infty d\omega W_{{\bf p}{\bf q},{\bf p}'{\bf q}'}^{cc'} \delta(\omega - \epsilon_{c,{\bf p'}} + \epsilon_{c,{\bf p}}) \delta(\omega - \epsilon_{c',{\bf q}} + \epsilon_{c',{\bf q'}})\delta ({\bf p}+{\bf q}-{\bf p}'-{\bf q}')
\nonumber\\
&~~~ \times n_{F}(\xi_{c,{\bf p}}) n_{F}(\xi_{c',{\bf q}}) [1 - n_{F}(\xi_{c,{\bf p}} + \omega)] [1 - n_{F}(\xi_{c',{\bf q}} - \omega)] 
\nonumber\\
&~~~ \times [\phi_c^{\alpha}(\xi_{c,{\bf p}})\boldsymbol{v}_{c,{\bf p}} + \phi_{c'}^{\alpha}(\xi_{c',{\bf q}})\boldsymbol{v}_{c',{\bf q}} - \phi_c^{\alpha}(\xi_{c,{\bf p}}+\omega)\boldsymbol{v}_{c,{\bf p'}} - \phi_{c'}^{\alpha}(\xi_{c',{\bf q}}- \omega) \boldsymbol{v}_{c',{\bf q'}}] \cdot {\bf F}_\alpha.
\end{align}
To make progress, it is convenient to separate the angular part of the momentum integrals as $\int_{\bf p'} = \int d\hat{\bf n}_{\bf p'} \int_{-\epsilon_{F}}^\infty d\xi_{1,{\bf p'}} \nu_1(\epsilon_{1,{\bf p'}})$ or $\int_{\bf p'} = \int d\hat{\bf n}_{\bf p'} \int_{-\infty}^{\Delta-\epsilon_{F}} d\xi_{2,{\bf p'}} \nu_2(\epsilon_{2,{\bf p'}})$. Here, we defined the unit vector $\hat{\bf n}_{\bf p'} = {\bf p'} / |{\bf p'}|$, normalized the angular integrals as $\int d\hat{\bf n}_{\bf p'} = 1$, and introduced the density of states for the two bands as $\nu_1(\epsilon_{1,{\bf p}}) = (2m_1)^{3/2} \sqrt{\epsilon_{1,{\bf p}}} / (4\pi^2) = m_1p/(2\pi^2)$ and $\nu_2(\epsilon_{2,{\bf p}}) = (2m_2)^{3/2} \sqrt{\Delta - \epsilon_{2,{\bf p}}} / (4\pi^2) = m_2p/(2\pi^2)$. Now, the delta functions containing $\omega$ can be used to perform the integrations in $\xi_{c,{\bf p'}}$ and $\xi_{c',{\bf q'}}$. 

Our aim is to extract the leading dependence on $T$ and $\xi_{c,{\bf p}}$.  With this goal in mind, the density of states, velocities and interaction matrix element may be evaluated on the Fermi surface, and we can approximate $\delta({\bf p}+{\bf q}-{\bf p'}-{\bf q'})\approx p^{-3}_F \delta(\hat{\bf n}_{\bf p} + \hat{\bf n}_{\bf q} - \hat{\bf n}_{\bf p'} - \hat{\bf n}_{\bf q'})$. The result of these transformations is
\begin{align}
I_{ee}^{cc'}\{\phi\}
& = - \frac{\nu_c(\epsilon_{F}) [\nu_{c'}(\epsilon_{\rm F})]^2}{T} \sum_{\alpha\in \{E,T\}}\frac{F_\alpha p_{F}}{p_{F}^3} \int_{-\infty}^\infty d\xi_{c',{\bf q}} \int_{-\infty}^\infty d\omega n_{F}(\xi_{c,{\bf p}}) n_{F}(\xi_{c',{\bf q}})
\nonumber\\
& ~~~ \times [1 - n_{F}(\xi_{c,{\bf p}} + \omega)] [1 - n_{F}(\xi_{c',{\bf q}} - \omega)] {\bf \Psi}_\alpha(\hat{\bf n}_{\bf p};\xi_{c,{\bf p}},\xi_{c',{\bf q}},\omega) \cdot \hat{{\bf n}}_{{\bf F}_{\alpha}},\label{eq:intermediate}
\end{align}
where we defined ${\bf F}_\alpha = F_\alpha \hat{\bf n}_{{\bf F}_{\alpha}}$, and the lower (upper) integration limit in $\xi_{1,{\bf q}}$ ($\xi_{2,{\bf q}}$) was extended to $\mp\infty$. We also introduced the angular integral
\begin{align}
{\bf \Psi}_\alpha(\hat{\bf n}_{\bf p};\xi_{c,{\bf p}},\xi_{c',{\bf q}},\omega)
& = \int d\hat{{\bf n}}_{\bf q} d\hat{{\bf n}}_{\bf p'}d\hat{{\bf n}}_{\bf q'} \tilde{W}_{\hat{{\bf n}}_{\bf p}\hat{{\bf n}}_{\bf q}\hat{{\bf n}}_{\bf p'}\hat{{\bf n}}_{\bf q'}}^{cc'} \delta(\hat{{\bf n}}_{\bf p}+\hat{{\bf n}}_{\bf q}-\hat{{\bf n}}_{\bf p'}-\hat{{\bf n}}_{\bf q'})
\nonumber\\
& ~~~ \times [\chi_c \phi_c^\alpha({\xi}_{c,{\bf p}})\hat{{\bf n}}_{\bf p} + \chi_{c'} \phi_{c'}^\alpha({\xi}_{c',{\bf q}})\hat{{\bf n}}_{\bf q} - \chi_c \phi_c^\alpha(\xi_{c,{\bf p}} + \omega)\hat{{\bf n}}_{\bf p'} - \chi_{c'} \phi_{c'}^\alpha(\xi_{c',{\bf q}} - \omega) \hat{{\bf n}}_{\bf q'}],
\end{align}
where $\chi_1 = 1/m_1$, $\chi_2 = -1/m_2$, and $\tilde{W}_{\hat{{\bf n}}_{\bf p}\hat{{\bf n}}_{\bf q}\hat{{\bf n}}_{\bf p'}\hat{{\bf n}}_{\bf q'}}^{cc'} = \tilde{W}_{cc'}(\theta,\varphi)$ is obtained from $W$ by fixing all incoming and outgoing momenta to $p_{F}$. We use two angles for parametrizing the matrix element $\tilde{W}_{cc'}$: the angle $\theta$ between the two incoming momenta ${\bf p}$ and ${\bf q}$, and the angle $\varphi$ between the planes spanned by ${\bf p}$ and ${\bf q}$ and by ${\bf p'}$ and ${\bf q'}$, respectively. With these conventions, the momentum transfer during collisions can be expressed as $k = |{\bf p}-{\bf p}'| = 2p_{F} \sin(\theta/2) \cos(\varphi/2)$.

For the integration in $\hat{{\bf n}}_{\bf p'}$ and $\hat{{\bf n}}_{\bf q'}$ we use a coordinate system in which the $z$ axis aligned with $\hat{{\bf n}}_{\bf p}+\hat{{\bf n}}_{\bf q}=\hat{{\bf n}}_{\bf p'}+\hat{{\bf n}}_{\bf q'}$, and employ the relation
\begin{align}
\delta(\hat{{\bf n}}_{\bf p}+\hat{{\bf n}}_{\bf q}-\hat{{\bf n}}_{\bf p'}-\hat{{\bf n}}_{\bf q'})
= \frac{\delta(\theta_{\bf p'}-\theta_{\bf q'})\delta(\varphi_{\bf q'}-\varphi_{\bf p'}-\pi)\delta(\theta_{\bf p'}-\theta/2)}{2\cos(\theta/2) \sin^2(\theta/2)}.
\end{align}
Next, we set $\varphi_{{\bf p}'} = \varphi$ and choose new coordinates so that the $z$-axis points along $\hat{\bf n}_{\bf p}$ and $\theta_{\bf q}=\theta$. The unit vectors $\hat{\bf n}_{\bf q}$, $\hat{\bf n}_{\bf p'}$ and $\hat{\bf n}_{\bf q'}$ can now be averaged in $\varphi_{\bf q}$. Denoting $\bar{\bf n}_{\bf k} = \int d(\varphi_{\bf q}/2\pi)\hat{{\bf n}}_{\bf k}$, the angular integral takes the form
\begin{align}
{\mathbf \Psi}_\alpha(\hat{{\bf n}}_{\bf p};\xi_{c,{\bf p}},\xi_{c',{\bf q}},\omega)
& = \frac{\pi}{(4\pi)^3} \int_0^\pi \frac{\sin\theta d\theta}{\cos(\theta/2)} \int_0^{2\pi} d\varphi \tilde{W}_{cc'}(\theta,\varphi)
\nonumber\\
& ~~~ \times [\chi_c \phi_c^\alpha({\xi}_{c,{\bf p}})\bar{\bf n}_{\bf p} + \chi_{c'} \phi_{c'}^\alpha({\xi}_{c',{\bf q}})\bar{\bf n}_{\bf q} - \chi_c \phi_c^\alpha(\xi_{c,{\bf p}} + \omega)\bar{\bf n}_{\bf p'} - \chi_{c'} \phi_{c'}^\alpha(\xi_{c',{\bf q}} - \omega) \bar{\bf n}_{\bf q'}],
\label{eq:Psi}
\end{align}
with $\bar{\bf n}_{\bf p} = \hat{\bf n}_{\bf p}$, $\bar{\bf n}_{\bf q} = \cos\theta~\hat{\bf n}_{\bf p}$, and $\bar{\bf n}_{{\bf p'}/{\bf q'}} = [\cos^2(\theta/2) \mp \cos\varphi\sin^2(\theta/2)] \hat{\bf n}_{\bf p}$. 

Turning our attention to the remaining integrations in $\omega$ and $\xi_{\bf q}$, a further simplification may be achieved with the help of the identity
\begin{align}
\int_{-\infty}^\infty d\xi_{c',{\bf q}} \int_{-\infty}^\infty d\omega n_{F}(\xi_{c',{\bf q}}) [1 - n_{F}(\xi_{c,{\bf p}} + \omega)]  [1 - n_{F}(\xi_{c',{\bf q}} - \omega)] \phi^{\alpha}(x_j)
= \int d\omega \omega n_{B}(\omega) [1 - n_{F}(\xi_{c,{\bf p}}+\omega)] \phi^{\alpha}(y_j),
\label{eq:enint}
\end{align}
for $(x_1, x_2, x_3, x_4) = (\xi_{c,{\bf p}}, \xi_{c',{\bf q}}, \xi_{c,{\bf p}} + \omega, \xi_{c',{\bf q}} - \omega)$ and $(y_1, y_2,$ $y_3, y_4) = (\xi_{c,{\bf p}}, - \xi_{c,{\bf p}} - \omega, \xi_{c,{\bf p}} + \omega, \xi_{c,{\bf p}} + \omega)$. This result was obtained with the help of the two relations $n_{F}(-\xi) = 1 - n_{F}(\xi)$ and $\int d\xi n_{F}(\xi) [1 - n_{F}(\xi-\omega)] = \omega n_{B}(\omega)$ and by relabeling the integration variables whenever convenient. After combining Eq.~\eqref{eq:intermediate} with Eqs.~\eqref{eq:Psi} and \eqref{eq:enint}, one arrives at the form of the collision integral stated in Eq.~\eqref{Coll_Int_ee_offdiag} in the main text.

\section{Solving the Boltzmann equation in the Fermi-liquid approximation}
\label{Appendix:SolvingBoltzmannEquation}

In this Appendix, we describe how the coupled set of linearized Boltzmann equations in \eqref{BoltzmannEq_Linearized} can be solved when the electron-electron collision integrals are treated in the Fermi-liquid approximation. Presenting the non-equilibrium part of the distribution function $\delta f_{c,{\bf p}}$  in the form stated in connection with Eq.~\eqref{Coll_Int_ee_2}, and the electron-electron collision integral as in Eq.~\eqref{Coll_Int_ee_tot}, one can present the the linearized Boltzmann equation in the form

\begin{align}
\left(
\begin{array}{c}
1
\\
\beta \xi_{c,{\bf p}}
\end{array}
\right)
= \frac{1}{\tau_{ei,c}} 
\left(
\begin{array}{c}
\phi^{E}_{c,s}(\xi_{c,{\bf p}})
\\
\phi^{T}_{c,a}(\xi_{c,{\bf p}})
\end{array}
\right) 
+ \frac{4}{\pi^2T^2} \frac{1}{\tau_{out,c}} \int_{-\infty}^\infty d\omega K(\omega,\xi_{c,{\bf p}}) \left(
\begin{array}{c}
\phi^{E}_{c,s}(\xi_{c,{\bf p}}) - \Lambda_{c,s} \phi^{E}_{c,s}(\xi_{c,{\bf p}} + \omega) + \Xi_{c,s} \phi^{E}_{\bar{c},s}(\xi_{c,{\bf p}}+\omega)
\\
\phi^{T}_{c,a}(\xi_{c,{\bf p}}) - \Lambda_{c,a} \phi^{T}_{c,a}(\xi_{c,{\bf p}} + \omega) + \Xi_{c,a} \phi^{T}_{\bar{c},a}(\xi_{c,{\bf p}}+\omega)
\end{array}
\right),
\label{BoltzmannEq_Linearized_Refined_2}
\end{align}
whereas $\phi_{c,a}^E$ and $\phi_{c,s}^T$ vanish. Eq.~\eqref{BoltzmannEq_Linearized_Refined_2} can be brought to a standard form~\cite{Jensen68,Sykes70}:
\begin{align}
\mathbb{X}(x)
= (x^2 + \pi^2 \varepsilon_c^2) \mathbb{Q}_c(x) - \int_{-\infty}^{\infty} dy (y - x)~{\rm csch}\bigg(\frac{y - x}{2}\bigg) \big[ \mathbb{M}_c \mathbb{Q}_c(y) - \mathbb{W}_c \mathbb{Q}_{\bar{c}}(y) \big].
\label{BoltzmannEq_Linearized_Refined_3}
\end{align}
Here, we define the two-component functions
\begin{align}
\mathbb{X}(x) 
& = \left(
\begin{array}{c}
1
\\
x
\end{array}
\right) {\rm sech}\bigg(\frac{x}{2}\bigg),
\label{Function_X}
\\
\mathbb{Q}_c(x) 
& = \left(
\begin{array}{c}
\hat{Q}_{c,s}^E(x)
\\
\hat{Q}_{c,a}^T(x)
\end{array}
\right)
= \frac{2}{\pi^2} \frac{1}{\tau_{out,c}} \left(
\begin{array}{c}
\hat{\phi}_{c,s}^E(x)
\\
\hat{\phi}_{c,a}^T(x)
\end{array}
\right) {\rm sech}\bigg(\frac{x}{2}\bigg),
\label{Function_Q}
\end{align}
with dimensionless variables $x = \beta\xi_{c,{\bf p}}$ and $y = \beta\omega$, and the diagonal matrices $\mathbb{M}_c = \mbox{diag}(\Lambda_{c,s},\Lambda_{c,a})$ and $\mathbb{W}_c = (\tau_{out,\bar{c}}/\tau_{out,c}) \mbox{diag}(\Xi_{c,s},\Xi_{c,a})$, 
and introduce the parameter $
\varepsilon_c = \sqrt{1 + {\tau_{out,c}}/{2\tau_{ei,c}}}$. We also defined $\hat{Y}(x) = Y(x/\beta) = Y(\xi_{\bf p})$ for $Y\in\{Q,\phi\}$. The integral equation \eqref{BoltzmannEq_Linearized_Refined_3}  can be converted to an inhomogeneous second order differential equation with the help of the Fourier transformation $
\tilde{\mathbb{Q}}_c(\tilde{x}) 
= \int_{-\infty}^\infty dx e^{i\tilde{x}x} \mathbb{Q}_c(x)$: 
 \begin{align}
\bigg[ \mathbb{I} \frac{d^2}{d\tilde{x}^2} + \pi^2 \big\{ 2~{\rm sech}^2(\pi \tilde{x}) \mathbb{M}_c - \varepsilon_c^2 \mathbb{I} \big\} \bigg] \tilde{\mathbb{Q}}_c(\tilde{x}) - 2\pi^2{\rm sech}^2(\pi \tilde{x}) \mathbb{W}_c \tilde{\mathbb{Q}}_{\bar{c}}(\tilde{x}) 
= - \tilde{\mathbb{X}}(\tilde{x}),
\label{BoltzmannEq_Linearized_Refined_4}
\end{align}
where $
\tilde{\mathbb{X}}(\tilde{x})= 2\pi~{\rm sech}(\pi \tilde{x}) (
1,i\pi~{\rm tanh}(\pi \tilde{x}))^t$, and $\mathbb{I}$ is the identity matrix. 

We first consider the homogeneous part of the equation and temporarily neglect the influence of interband collisions, $\mathbb{W}_c = \tilde{\mathbb{X}} = 0$ in Eq.~\eqref{BoltzmannEq_Linearized_Refined_4}. Following the solution strategy proposed in Refs.~\cite{Jensen68,Brooker68},  we equip the homogeneous equation with an additional parameter $\lambda$ as
\begin{align}
\bigg[ \frac{d^2}{d\tilde{x}^2} + \pi^2 \big\{ 2\lambda~{\rm sech}^2(\pi \tilde{x}) - \varepsilon_c^2 \big\} \bigg] \tilde{Q}_c(\tilde{x}) = 0.
\label{KineticEq_NoMomentum_Homogeneous}
\end{align}
This equation is similar in structure to a time-independent Schr\"{o}dinger equation in a ${\rm sech}^2 x$ potential~\cite{LL:QM}. 

For a particular choice of eigenvalues, $
\lambda_{cn}(\varepsilon_c) = (n + \varepsilon_c) (n + \varepsilon_c + 1)/2$, Eq.~\eqref{KineticEq_NoMomentum_Homogeneous} is solved by the eigenfunctions
\begin{align}
\tilde{Q}_{cn}(\tilde{x})
= [{\rm sech}(\pi \tilde{x})]^{\varepsilon_c}~_2 F_1[-n, n + 2\varepsilon_c + 1, \varepsilon_c + 1, (1 - {\rm tanh}(\pi \tilde{x}))/2]
\label{Eigenfunction_NoMomentum}
\end{align}
for $n \in \{0, \mathbb{N}\}$. Here, $_2 F_1$ is the hypergeometric function. 

The eigenfunctions are even(odd)-symmetric in $\tilde{x}$ for even (odd) integer $n$, and satisfy the orthogonal relation
\begin{align}
\int_{-\infty}^\infty d\tilde{x}~{\rm sech}^2(\pi \tilde{x}) [\tilde{Q}_{cm}(\tilde{x})]^* \tilde{Q}_{cn}(\tilde{x})
= \frac{n! 2^{2\varepsilon_c + 1} [\Gamma(\varepsilon_c + 1)]^2 \delta_{mn}}{\pi (2n + 2\varepsilon_c + 1) \Gamma(n + 2\varepsilon_c + 1)},
\label{Orthogonality}
\end{align}
where $\Gamma(z)$ is the Gamma function, and $\delta_{mn}$ is the Kronecker delta. 

We now derive the solution for Eq.~\eqref{BoltzmannEq_Linearized_Refined_4}. Let us expand $\tilde{\mathbb{Q}}_c(\tilde{x})$ in eigenfunctions basis with fixed index $c'=$1 or 2
\begin{align}
\tilde{Q}_{c,s}^E(\tilde{x})
& = \sum_{n=0}^\infty C_{cc',2n}^E \tilde{Q}_{c',2n}(\tilde{x}),
\label{SeriesSolution_E}
\\
\tilde{Q}_{c,a}^T(\tilde{x})
& = \sum_{n=0}^\infty C_{cc',2n+1}^T \tilde{Q}_{c',2n+1}(\tilde{x}),
\label{SeriesSolution_T}
\end{align}
with the expansion coefficients $C_{cc',n}^\alpha$. Plugging these series expansions in Eq.~\eqref{BoltzmannEq_Linearized_Refined_4}, and using Eqs.~\eqref{KineticEq_NoMomentum_Homogeneous} and \eqref{Orthogonality}, we derive coupled algebraic equations for $C_{cn}^\alpha$
\begin{align}
\left(
\begin{array}{cc}
\lambda_{1,2n} - \Lambda_{1,s} & \Xi_{1,s} \tau_{out,2}/\tau_{out,1}
\label{CoupledEq_E}
\\
\Xi_{2,s} \tau_{out,1}/\tau_{out,2} & \lambda_{2,2n} - \Lambda_{2,s}
\end{array}
\right)
\left(
\begin{array}{c}
C_{1c',2n}^E
\\
C_{2c',2n}^E
\end{array}
\right)
& = \left(
\begin{array}{c}
1
\\
1
\end{array}
\right) D_{c',2n}^E,
\\
\left(
\begin{array}{cc}
\lambda_{1,2n+1} - \Lambda_{1,a} & \Xi_{1,a} \tau_{out,2}/\tau_{out,1}
\\
\Xi_{2,a} \tau_{out,1}/\tau_{out,2} & \lambda_{2,2n+1} - \Lambda_{2,a}
\end{array}
\right)
\left(
\begin{array}{c}
C_{1c',2n+1}^T
\\
C_{2c',2n+1}^T
\end{array}
\right)
& = \left(
\begin{array}{c}
1
\\
1
\end{array}
\right) D_{c',2n+1}^T,
\label{CoupledEq_T}
\end{align}
Here, we defined
\begin{align}
D_{c,2n}^E
& = \frac{\int_{-\infty}^\infty d\tilde{x}~{\rm sech}(\pi \tilde{x}) [\tilde{Q}_{c,2n}(\tilde{x})]^*}{\pi \int_{-\infty}^\infty d\tilde{x}~{\rm sech}^2(\pi \tilde{x}) |\tilde{Q}_{c,2n}(\tilde{x})|^2}=\frac{(2n + \varepsilon_c + 1/2) \Gamma(n+\varepsilon_c+1/2)\Gamma(n+(\varepsilon_c+1)/2)}{\pi\Gamma(n+\varepsilon_c/2+1)\Gamma(n+1)\Gamma(\varepsilon_c+1)},\label{Coefficient_D_E}
\\
D_{c,2n+1}^T
& = \frac{i \int_{-\infty}^\infty d\tilde{x}~{\rm sech}(\pi \tilde{x}) {\rm tanh}(\pi \tilde{x}) [\tilde{Q}_{c,2n+1}(\tilde{x})]^*}{\int_{-\infty}^\infty d\tilde{x}~{\rm sech}^2(\pi \tilde{x}) |\tilde{Q}_{c,2n+1}(\tilde{x})|^2}=\frac{i (2n + \varepsilon_c + 3/2) \Gamma(n+\varepsilon_c+3/2)\Gamma(n+(\varepsilon_c+1)/2)}{\Gamma(n+\varepsilon_c/2+2)\Gamma(n+1)\Gamma(\varepsilon_c+1)}.\label{Coefficient_D_T}
\end{align}
After performing the matrix inversion in Eqs.~\eqref{CoupledEq_E} and \eqref{CoupledEq_T}, we find the expansion coefficients
\begin{align}
C_{cc,2n}^E
& = \frac{(\lambda_{\bar{c},2n} - \Lambda_{\bar{c},s} - \Xi_{c,s} \tau_{out,\bar{c}}/\tau_{out,c}) D_{c,2n}^E}{(\lambda_{c,2n} - \Lambda_{c,s})(\lambda_{\bar{c},2n} - \Lambda_{\bar{c},s}) - \Xi_{c,s} \Xi_{\bar{c},s}},
\label{Coefficient_C_E}
\\
C_{cc,2n+1}^T
& = \frac{(\lambda_{\bar{c},2n+1} - \Lambda_{\bar{c},a} - \Xi_{c,a} \tau_{out,\bar{c}}/\tau_{out,c}) D_{c,2n+1}^T}{(\lambda_{c,2n+1} - \Lambda_{c,a})(\lambda_{\bar{c},2n+1} - \Lambda_{\bar{c},a}) - \Xi_{c,a} \Xi_{\bar{c},a}}.
\label{Coefficient_C_T}
\end{align}
In the clean limit of $\varepsilon_c = 1$, the coefficients are reduced to a form consistent with Ref.~\cite{Li18}.

\section{Conductivities}
\label{Appendix:Conductivities}

In view of Eq.~\eqref{CurrentDensity}, the electric and thermal conductivities can be expressed in terms of the functions $\hat{Q}_{c,s}^E(x)$ and $\hat{Q}_{c,a}^T(x)$ as 
\begin{align}
\left(
\begin{array}{c}
L_{EE}
\\
L_{TT}
\end{array}
\right)
= \frac{\pi^2 \mathcal{N}}{2} \sum_{c\in\{1,2\}} \frac{\tau_{out,c}}{m_c} \int_{-\infty}^{\infty} dx \frac{\partial n_{F}(x)}{\partial x} {\rm cosh}\bigg(\frac{x}{2}\bigg) \left(
\begin{array}{c}
- e^2 \hat{Q}_{c,s}^E(x)
\\
- T x \hat{Q}_{c,a}^T(x)
\end{array}
\right).
\label{ElectricCond_2}
\end{align}
For the purpose of this calculation, it is then convenient to transform to the conjugate variable $\tilde{x}$, 
\begin{align}
\left(
\begin{array}{c}
L_{EE}
\\
L_{TT}
\end{array}
\right)
= \frac{\pi^2 \mathcal{N}}{8} \sum_{c\in\{1,2\}} \frac{\tau_{out,c}}{m_c} \int_{-\infty}^\infty d\tilde{x}~{\rm sech}(\pi \tilde{x}) \left(
\begin{array}{c}
e^2 \tilde{Q}_{c,s}^E(\tilde{x})
\\
- i\pi T~{\rm tanh}(\pi \tilde{x}) \tilde{Q}_{c,a}^T(\tilde{x})
\end{array}
\right).
\label{ElectricCond_3}
\end{align}
Inserting Eq.~\eqref{SeriesSolution_E} in Eq.~\eqref{ElectricCond_3}, one obtains a series expansion for the electric conductivity $\sigma\equiv L_{EE}$,
\begin{align}
\sigma
= \frac{\pi^2}{8} \mathcal{N}e^2 \sum_{c\in\{1,2\}} \frac{\tau_{out,c}}{m_c} \sum_{n=0}^\infty C_{cc,2n}^E \int_{-\infty}^\infty d\tilde{x}~{\rm sech}(\pi \tilde{x}) \tilde{Q}_{c,2n}(\tilde{x}).
\label{ElectricCond_5}
\end{align}
For the thermal conductivity, we insert Eq.~\eqref{SeriesSolution_T} in Eq.~\eqref{ElectricCond_3} and find $\kappa\approx L_{TT}$ as \begin{align}
\kappa
= - i \frac{\pi^3}{8} \mathcal{N} T \sum_{c\in\{1,2\}} \frac{\tau_{out,c}}{m_c} \sum_{n=0}^\infty C_{cc,2n+1}^T \int_{-\infty}^\infty d\tilde{x}~{\rm sech}(\pi \tilde{x}) {\rm tanh}(\pi \tilde{x}) \tilde{Q}_{c,2n+1}(\tilde{x}).
\end{align}

\end{widetext}

\section{Ansatz of constant $\phi_c^E$ and $\phi_c^T/\xi_{c,{\bf p}}$}
\label{app:ansatz}

Here we discuss, how one may directly obtain electric and thermal conductivities in the form of Eqs.~\eqref{eq:sigmaGRTA} and \eqref{eq:kappaGRTA} from the coupled kinetic equations \eqref{BoltzmannEq_Linearized}, under certain simplifying assumptions. The goal here is to further motivate the phenomenological expressions, in particular the ansatz of unequal $\tau_E$ and $\tau_T$. To this end, we analyze the forward-scattering dominated regime as an instructive example. We assume that typical transferred momenta $k$ during collisions are small, $k\ll p_F$. This is relevant, for example, for the statically screened Coulomb interaction, when the Thomas-Fermi screening wave number $k_{TF}$ fulfills $k_{TF}\ll p_F$. It has been argued \cite{Maslov11,Li18} that the ansatz of a $\xi$-independent $\phi^E$ is well suited for the forward scattering limit, because ``vertical" relaxation to the Fermi surface is the fastest process. One may expect that in a disordered compensated metal this ansatz can work particularly well, since in the absence of interactions, when impurities provide the only scattering mechanism, the Boltzmann equation is solved by a constant $\phi^E_c=\tau_{ei,c}$. 

For the electric conductivity, assuming a constant $\phi_E$, the two coupled Boltzmann equations in the forward scattering dominant regime can be simplified considerably,
\begin{align}
&e\boldsymbol{v}_{c,{\bf p}}\cdot {\bf E} \frac{\partial n_{F}(\xi_{c,{\bf p}})}{\partial \xi_{c,{\bf p}}}\label{eq:phiEeq}\\=&e\boldsymbol{v}_{c,{\bf p}}\cdot{\bf E}\frac{\partial n_{F}(\xi_{c,{\bf p}})}{\partial \xi_{c,{\bf p}}}\left[\frac{\phi^E_{c}}{\tau_{ei,c}}+m_c\left(\frac{\phi^E_1}{m_1}+\frac{\phi^E_2}{m_2}\right)\Gamma^{c\bar{c}}_p\right],\nonumber
\end{align}
with 
\begin{align}
\Gamma^{c\bar{c}}_p=\frac{m_cm_{\bar{c}}^2}{2^9\pi^7 p^3} [\xi_{c,{\bf p}}^2+(\pi T)^2]\int dk k^2 W^{c\bar{c}}(k),
\label{eq:Gammas}
\end{align}
$W^{12}=W^{21}$ and $m_1\Gamma^{12}_p=m_2\Gamma^{21}_p$. As expected, the intraband collision integral does not give a contribution to this equation. 

Equation~\eqref{eq:phiEeq} does not have a solution for a constant $\phi^E$, since $\Gamma^{c\bar{c}}_p$ is momentum-dependent. One may think of fixing $\Gamma^{c\bar{c}}_p$ to the Fermi surface, but this step would completely neglect the influence of the $\xi^2$ term in Eq.~\eqref{eq:Gammas}, which tends to increase the scattering rate. Here, we will follow an alternative approach \cite{Appel78,Pal12,Li18}, multiply Eq.~\eqref{eq:phiEeq} by $\boldsymbol{v}_{c,{\bf p}}$ and integrate in ${\bf p}$. As a result, a new momentum-independent scattering rate $1/\tau_{E,ee}^{c\bar{c}}=\frac{4}{3}\Gamma^{c\bar{c}}_{p_F}$ appears, and the equation for $\phi^E_c$ reads 
\begin{align}
\left(\begin{array}{cc} \frac{m_1}{\tau_{ei,1}}+\frac{m_1}{\tau_{E,ee}^{12}}&\frac{m_1}{\tau^{12}_{E,ee}}\\
\frac{m_2}{\tau^{21}_{E,ee}}&\frac{m_2}{\tau_{ei,2}}+\frac{m_2}{\tau_{E,ee}^{21}}\end{array}\right)\left(\begin{array}{cc}\frac{\phi^E_1}{m_1}\\\frac{\phi^E_2}{m_2}\end{array}\right)=\left(\begin{array}{cc} 1\\1\end{array}\right).
\label{eq:PhEforward}
\end{align}
This equation is structurally similar to Eq.~\eqref{eq:vKeyes} (limited to the case of an applied electric field). After solving Eq.~\eqref{eq:PhEforward} the conductivity can be found with the help of Eq.~\eqref{CurrentDensity}. Defining $\tau_E$ through the relation $m_c\Gamma^{c\bar{c}}=\tilde{m}/\tau_E$, and $\tau_{el}$ as in \eqref{eq:tauel}, one obtains the electric conductivity in the form of Eq.~\eqref{eq:sigmaGRTA}. 

For the calculation of the thermal conductivity we use the ansatz $\phi^T_c(\xi_{c,{\bf p}})=\varphi_c^T\xi_{c,{\bf p}}/T$ with constant $\varphi_c^T$. With this ansatz, one obtains two {\it decoupled} equations for $\varphi_1^T$ and $\varphi_2^T$,
\begin{align}
&\xi_{c,{\bf p}}\boldsymbol{v}_{c,{\bf p}}\cdot \nabla T\frac{\partial n_{F}(\xi_{c,{\bf p}})}{\partial \xi_{c,{\bf p}}}\label{eq:phiTforward}\\
=&\varphi_{c}^T\xi_{c,{\bf p}}\boldsymbol{v}_{c,{\bf p}}\cdot \nabla T\frac{\partial n_{F}(\xi_{c,{\bf p}})}{\partial \xi_{c,{\bf p}}}\left[\frac{1}{\tau_{c,ei}}+{R}^{cc}_p+{R}^{c\bar{c}}_p\right],\nonumber
\end{align}
where 
\begin{align}
{R}^{cc'}_p=\frac{1}{3}\frac{m_cm_{c'}^2}{2^7\pi^7 p} [\xi_{c,{\bf p}}^2+(\pi T)^2]\int dk W^{cc'}(k).
\end{align}
The key difference in the expression for the rate $R$ compared to $\Gamma$ is the lack of the factor $k^2$ under the integral \cite{Li18}. Small-$k$ scattering processes are much less effective for the electric conductivity compared to the thermal conductivity. In analogy to the electric field case, we multiply Eq.~\eqref{eq:phiTforward} by $\xi_{c,{\bf p}}\boldsymbol{v}_{c,{\bf p}}\nabla T$, and integrate over momenta. We find that the relevant temperature-dependent scattering rate becomes $1/\tau_{T,ee}^{cc'}=\frac{12}{5} R^{cc'}_{p_F}$ \cite{Remark5}. In view of the relation $m_1/\tau^{12}_{T,ee}=m_2/\tau^{21}_{T,ee}$, and motivated by the results obtained for the RTA in Sec.~\ref{sec:RTA}, we define a new time scale $\tau_T$ by setting $\tilde{m}/\tau_{T}=m_c/\tau_{T,ee}^{c\bar{c}}$. This leads to $\varphi_c^T=\tilde{\tau}_c$ [compare Eq.~\eqref{eq:taucGTRA}] and the thermal conductivity takes the form of  Eq.~\eqref{eq:kappaGRTA}. 

Relations analogous to \eqref{eq:sigmaGRTA} and \eqref{eq:kappaGRTA} have previously been obtained in Ref.~\cite{Li18} for the clean case. While the result for the electric conductivity is consistent with Eq.~\eqref{eq:sforward} when applied to the screened Coulomb interaction, the result for the thermal conductivity obtained with the ansatz of constant $\varphi^T_c$ differs from Eq.~\eqref{eq:kappaforward} as it predicts a prefactor $10/(3\pi^2)$ compared to $4/\pi^2$ in Eq.~\eqref{eq:sforward}, previously noted in Ref.~\cite{Li18}. This makes obvious the necessity for a nonperturbative solution for $\varphi^T_c$, even after restricting considerations to the forward scattering regime and clean systems. Despite this shortcoming, the considerations presented in this Appendix clearly demonstrate the importance of introducing two different time scales, $\tau_E$ and $\tau_T$, instead of the single scale $\tau$ considered in Sec.~\ref{sec:RTA}.

\section{Fixing $\tau_{ee}$ - Boundary condition in the clean limit}
\label{Appendix:BC}

The relation between $\tau_{ee}$ and $\tau_{out}$ can be fixed by using the boundary condition for both electric and thermal conductivities in the clean limit. 

The electric and thermal conductivities in the RTA in the clean limit read as
\begin{align}
\sigma&=\frac{\mathcal{N}e^2\tau_E}{\tilde{m}},
\label{ElectCond_RTA_Clean}\\
\kappa
& = \frac{\pi^2}{3} \mathcal{N}T \sum_{c\in\{1,2\}} \frac{\tilde{\tau}^c_{T,ee}}{m_c},
\label{ThermCond_RTA_Clean}
\end{align}
where we defined $1/\tilde{\tau}^c_{T,ee} = 1/\tau_{T,ee}^{cc} + \tilde{m} / (m_c \tau_T)$. To arrive at Eq.~\eqref{ThermCond_RTA_Clean}, we approximated
$\langle\!\langle \xi_{c,{\bf p}}^2 \rangle\!\rangle \approx \pi^2 T^2 / 3$ and $\langle\!\langle \xi_{c,{\bf p}} \rangle\!\rangle \approx \pi^2 T^2 / (3\epsilon_F)$ in Eq.~\eqref{ThermalConductivity} for $T\ll\epsilon_F$ and consistently neglected the small quantity $\langle\!\langle \xi_{c,{\bf p}} \rangle\!\rangle^2$ compared to $\langle\!\langle \xi_{c,{\bf p}}^2 \rangle\!\rangle$.

\subsection{Hubbard interaction}

For the Fermi-liquid solution Eq.~\eqref{ElectricCond_General}, we take the clean limit, and additionally neglect intraband scattering, since we expect its influence on the electric conductivity to be small. We can use Eq.~\eqref{ElectricCond_Inter} and perform the summation numerically for $\tilde{\Lambda}=1/3$. The result is $\sigma\approx A_{\sigma}\tilde{\sigma}_0$ with $A_\sigma=0.632$ and $\tilde{\sigma}_0=\mathcal{N}e^2\epsilon_F/(m_cu_{c\bar{c}}T^2)$. Comparing this result with Eq.~\eqref{ElectCond_RTA_Clean}, we obtain
\begin{align}
\frac{1}{\tau_E}=\frac{m_c}{\tilde{m}}\frac{u_{c\bar{c}}}{A_{\sigma}}\frac{T^2}{\epsilon_F}
\label{eq:Rtau_E}
\end{align}

A similar strategy can be used for the thermal conductivity. The Fermi-liquid solution Eq.~\eqref{ThermalCond_General} is approximated by
\begin{align}
\kappa
& \approx \frac{\pi^2}{8} \mathcal{N}T A_\kappa \sum_{c\in\{1,2\}} \frac{\tau_{out,c}}{m_c},
\label{ThermCond_FL_Clean}
\end{align}
where we define
\begin{align}
A_\kappa
& = \frac{3}{8}\sum_{n=0}^\infty 
\frac{4n + 5}{(n+1)(2n+3)[(n+1)(2n+3) - 1/3]}
\nonumber\\
& \approx 0.294.
\label{C_3}
\end{align}

If we compare Eqs.~\eqref{ThermCond_RTA_Clean} and \eqref{ThermCond_FL_Clean} term by term, we can find the relations
\begin{align}
\frac{1}{\tau_{T,ee}^{cc}}
& =  \frac{u_{cc}}{A_\kappa} \frac{T^2}{\epsilon_{F}},
\\
\frac{1}{\tau_T}
& =  \frac{m_c}{\tilde{m}} \frac{u_{c\bar{c}}}{A_\kappa} \frac{T^2}{\epsilon_{F}}.
\label{eq:Rtau_T}
\end{align}
Comparing Eqs.~\eqref{eq:Rtau_E} with \eqref{eq:Rtau_T}, we find ${\tau_T}^{-1}/{\tau_E}^{-1}\approx 2.15$. Interband scattering is more effective for thermal transport than for electric transport.

\subsection{Screened Coulomb interaction}
For the screened Coulomb interaction, we follow the same strategy. We restrict our attention to the forward scattering limit, for which simple transparent results can be derived.

For the electric conductivity, it is important to note that in the low-temperature limit it is sufficient to consider interband scattering only [this corresponds to neglecting the difference $\phi_{c,s}^E(\xi_{c,{\bf p}})-\phi^E_{c,s}(\xi_{c,{\bf p}}+\omega)$ in Eq.~\eqref{Coll_Int_ee_diag}, which gives contribution that is subleading in temperature. These terms were kept for the general derivation in order to treat electric and thermal conductivities analogously]. With this insight, we can use Eq.~\eqref{ElectricCond_Inter} for the calculation of the electric conductivity. The most divergent contribution in the forward scattering limit originates from the $n=0$ term in the sum, for which we can further approximate $\tilde{\Lambda}\approx 1-2\tilde{k}_{TF}^2$. For the purpose of later reference, we note that the electric conductivity can be written as 
\begin{align}
\sigma&=\frac{3\mathcal{N} e^2}{32\tilde{k}_{TF}^2}\sum_{c\in\{1,2\}}\frac{\tau_{out,c}}{{m_c}},\label{eq:sigmalimit}\\
\frac{1}{\tau_{out,c}}&=\frac{\pi^4 T^2}{2^5} \frac{m_cm_{\bar{c}}^2}{(m_1+m_2)^2}\frac{\tilde{k}_{TF}}{p_F^2}.
\end{align} 
For the sake of consistency, only the interband contribution to $1/\tau_{out,c}$ has been considered.

For the thermal conductivity in the forward scattering limit, both interband and intraband scattering contributions need to be accounted for. Here, it is sufficient to approximate $\Lambda_{c,a}\approx 1$. Using the relation
\begin{align}
\sum_{n=0}^\infty \frac{4n+5}{(n+1)(2n+3)[(n+1)(2n+3)-1]}=1,
\end{align}
one finds the result
\begin{align}
\kappa&=\frac{\pi^2 \mathcal{N} T}{8}\sum_{c\in \{1,2\}}\frac{\tau_{out,c}}{m_c}\label{eq:kappalimit}\\
\frac{1}{\tau_{out,c}}&=\frac{\pi^4 T^2}{2^5}m_c\frac{m_c^2+m_{\bar{c}}^2}{(m_c+m_{\bar{c}})^2}\frac{\tilde{k}_{TF}}{p_F^2}.
\end{align}
In the last formula, the contribution proportional to $m_c^3$ ($m_cm^2_{\bar{c}}$) originates from intraband (interband) collisions.

By comparing the results \eqref{eq:sigmalimit} to the RTA result \eqref{ElectCond_RTA_Clean} one finds
\begin{align}
\frac{1}{\tau_E}&=\frac{\pi^4}{12}\frac{m_2}{m_1+m_2}\tilde{k}_{TF}^3\frac{T^2}{\epsilon_F}
\end{align}
Similarly, by comparing \eqref{eq:kappalimit} to the RTA result  \eqref{ThermCond_RTA_Clean}, one finds
\begin{align}
\frac{1}{\tau_{T,ee}^{cc'}}&=\frac{\pi^4}{24}\frac{m_cm_{c'}^2}{m_1(m_1+m_2)^2}\tilde{k}_{TF}\frac{T^2}{\epsilon_F}.
\end{align}
The latter expression results in 
\begin{align}
\frac{1}{\tau_T}=\frac{\pi^4}{24}\frac{m_2}{m_1+m_2}\tilde{k}_{TF}\frac{T^2}{\epsilon_F},
\end{align}
so that $\tau_T^{-1}/\tau_E^{-1}=1/(2\tilde{k}_{TF}^2)\gg 1$.

\bibliography{Library,library-1}

\begin{thebibliography}{44}
\bibitem[{Rem({\natexlab{a}})}]{Remark6}
\bibinfo{note}{Lorenz ratios $\mathcal{L}<\mathcal{L}_0$ have also been
  observed in a number of non-Fermi liquid systems, e.g.,
  Refs.~\cite{Paglione06,Ronning06,Tanatar07,Mahajan13}. Upward deviations
  $\mathcal{L}>\mathcal{L}_0$ are also feasible. An increase of the Lorenz ratio
  has, for example, been predicted to occur in two-dimensional
  disordered electron liquids due to the long-range Coulomb interaction
  \cite{Livanov91,Raimondi04,Niven05,Catelani05,Michaeli09,Schwiete14a,Schwiete16a,Schwiete16b}.}
\bibitem[{Rem({\natexlab{c}})}]{Remark1}
\bibinfo{note}{The electron-electron collision integrals
  [Eqs.~\eqref{Coll_Int_ee_diag} and \eqref{Coll_Int_ee_offdiag}] are overall
  consistent with Ref.~\cite{Li18}. There are two remarks. First, the intraband
  contribution [Eq.~\eqref{Coll_Int_ee_diag}] is ignored in Ref.~\cite{Li18} by
  assuming that the Baber mechanism is predominant under the electric field.
  Second, the numerical factor in the collision integral is mismatched with
  Ref.~\cite{Li18} by $2\pi$. However, this does not matter if we focus on the
  evaluation of the Lorenz ratio.}
\bibitem[{Rem({\natexlab{d}})}]{Remark2}
\bibinfo{note}{The identity $\sum_{n=0}^\infty \Upsilon_{c,n}^\sigma (2n +
  \varepsilon_c + 1/2)/(\lambda_{c,2n} - 1) = 1$ confirms that the electric
  conductivity is insensitive to electron-electron collisions at intermediate
  temperatures.}
\bibitem[{Rem({\natexlab{e}})}]{Remark3}
\bibinfo{note}{Our result is slightly different from $\mathcal{L}/\mathcal{L}_0
  = 0.46$ in Sec. V. B. in Ref.~\cite{Li18}, where $\tilde{W}_{11} =
  \tilde{W}_{22} = 0$ is assumed only for the electric conductivity while it is
  not the case for the thermal conductivity.}
\bibitem[{Rem({\natexlab{f}})}]{RemarkPhonons}
\bibinfo{note}{Phenomenologically, the effect of electron-phonon scattering can
  be included into our formalism by adding a contribution to the scattering
  rate $\tau^{-1}_{ei,c}$. Here, electric and thermal transport need to be
  distinguished. Due to the predominance of small-angle scattering, the
  scattering on acoustic phonons is more effective for thermal transport, with
  a characteristic rate $\tau^{-1}\propto T^3$, than for electric transport,
  for which $\tau^{-1}\propto T^5$ \cite{Ziman01}.}
\bibitem[{Rem({\natexlab{g}})}]{Remark5}
\bibinfo{note}{It is known from the exact solution in the clean case that
  (unlike for electric conductivity) the obtained coefficient acquires
  additional corrections of the order of 20\% \cite{Li18}}.
\end{thebibliography}

\end{document}